\documentclass[%
a4paper,
superscriptaddress,
notitlepage,
bibnotes,
 amsmath,amssymb,
 aps,
10pt,twocolumn,
pre,
]{revtex4-1}

\usepackage{soul}

\usepackage{xcolor}
\colorlet{mylinkcolor}{blue!66!black!80}

\usepackage{mathtools}

\usepackage[colorlinks=true,linkcolor=mylinkcolor,citecolor=mylinkcolor,filecolor=cyan,urlcolor=mylinkcolor,breaklinks=true]{hyperref}

\usepackage[utf8]{inputenc}

\newcommand{\mat}[1]{\textbf{#1}}
\newcommand{\avg}[1]{\langle#1\rangle}

\newcommand{\del}{\partial}
\newcommand{\dd}{{\rm d}}

\newcommand{\bbb}{\boldsymbol{b}}
\newcommand{\bc}{\boldsymbol{c}}

\newcommand{\by}{\boldsymbol{y}}
\newcommand{\bz}{\boldsymbol{z}}

\newcommand{\xtp}{x_{t+\dd t}}
\newcommand{\ytp}{y_{t+\dd t}}
\newcommand{\rtp}{r_{t+\dd t}}
\newcommand{\mtp}{m_{t+\dd t}}
\newcommand{\xtraj}{\{x_{t'}\}_{t'\le t}}
\newcommand{\ytraj}{\{y_{t'}\}_{t'\le t}}
\newcommand{\rtraj}{\{r_{t'}\}_{t'\le t}}
\newcommand{\mtraj}{\{m_{t'}\}_{t'\le t}}

\newcommand{\rr}{{\rm r}}
\newcommand{\rmb}{{\rm b}}
\newcommand{\x}{{\rm x}}
\newcommand{\y}{{\rm y}}

\newcommand{\m}{\mathrm{m}}
\newcommand{\bSigma}{\boldsymbol{\Sigma}}

\newcommand{\T}{\top}

\begin{document}
\title{Sensory capacity: an information theoretical measure of the performance of a sensor}
\author{David Hartich}
\affiliation{%
II. Institut für Theoretische Physik, Universität Stuttgart, 70550 Stuttgart, Germany}
\author{Andre C. Barato}%
\affiliation{%
II. Institut für Theoretische Physik, Universität Stuttgart, 70550 Stuttgart, Germany}
\affiliation{%
Max Planck Institute for the Physics of Complex Systems, N\"othnizer Straße 38, 01187 Dresden, Germany}%
 \author{Udo Seifert}%
\affiliation{%
II. Institut für Theoretische Physik, Universität Stuttgart, 70550 Stuttgart, Germany}

\begin{abstract}
For a general sensory system following an external stochastic signal, we introduce the sensory capacity. This quantity 
characterizes the performance of a sensor: sensory capacity is maximal if the instantaneous state of the sensor has as 
much information about a signal as the whole time-series of the sensor. We show that adding a memory to the sensor 
increases the sensory capacity. This increase quantifies the improvement of the sensor with the addition of the  memory. 
Our results are obtained with the framework of stochastic thermodynamics of bipartite systems,
which allows for the definition of an efficiency that relates the rate with which the sensor learns about the signal with 
the energy dissipated by the sensor, which is given by the thermodynamic entropy production. 
We demonstrate a general tradeoff between sensory capacity and efficiency: if the sensory capacity is equal to its maximum 1, 
then the efficiency must be less than 1/2. As a physical realization of a sensor we consider a two component cellular network 
estimating a fluctuating external ligand concentration as signal. This model leads to coupled linear Langevin equations 
that allow us to obtain explicit analytical results.
\end{abstract}

\maketitle

\section{Introduction}

The relation between information and thermodynamics is a very active topic, as reviewed in \cite{parr15}.
Prominently, developments in this field lead to a better understanding of fundamental limits related to dissipation in a computer 
and of cellular information processing. Much of the renewed interest in this relation between information and thermodynamics
is associated with the fact that recent experiments with small systems verify fundamental relations like the 
Landauer limit for the erasure of a bit \cite{beru12,jun14} and the conversion of information into work \cite{toya10,kosk14,rold14}. Theoretical
advances in the field include second law inequalities and fluctuation relations containing an informational term
\cite{saga08,cao09,saga10,horo10,horo11,gran11,abre11,abre12,baue12,muna12,muna13,saga12a,ito13,horo13,stras13,baue14,sand14,prok14,prok15,rosi15,bech15,shir15,shir15a}, 
generalization of thermodynamics to include information reservoirs \cite{mand12,mand13,deff13,deff13a,bara13,bara14,bara14a,hopp14,merh15,boyd15arxiv}, stochastic thermodynamics of bipartite systems
\cite{alla09,bara13b,hart14,dian14,horo14,horo14a,horo15}, and the relation between dissipation and information in biological systems \cite{lan12,meht12,bara13a,palo13,skog13,lang14,gove14a,gove14,bara14b,sart14,hart15,bo15,ito15}.

A sensor that learns about (or ``measures'') an external stochastic signal constitutes a fundamental setup within thermodynamics of
information processing. In this case energy is dissipated and the sensor obtains information about the external signal, in contrast to  
a Maxwell's demon, which is another fundamental setup, where information is used to extract work.

General results for the thermodynamics of a sensor have been obtained by Still et al. \cite{stil12}. They have shown that an entropy characterizing 
how much information the sensor obtains about the external signal is bounded by the dissipated heat. Similarly, we have shown that an 
entropic rate, dubbed learning rate, is bounded by the thermodynamic entropy production in bipartite systems \cite{bara14b}, which allowed for the 
definition of a thermodynamic efficiency for models related to cellular information processing. 

In this paper, using bipartite Markov processes we introduce the sensory capacity, an informational efficacy parameter characterizing the performance of a sensor. 
This quantity is defined as the learning rate divided by the transfer entropy rate, where the latter quantifies how much information the full time series of the sensor has about the signal. 
Sensory capacity is positive and bounded by 1. The limit 1 is reached if the information contained in the instantaneous state of
the sensor equals the information contained in the whole time-series of the sensor, which is the maximum information the sensor can have about the signal. 

A bare sensor, i.e., a sensor with only one degree of freedom, is compared to a sensor that contains a memory, which is a second degree of freedom. We 
show that the addition of a memory to a bare sensor can increase the sensory capacity. This increase in sensory capacity quantifies how much of the information contained in the time-series 
of the bare sensor is stored in the instantaneous state of the memory. 

Our results are obtained with coupled linear Langevin equations that constitute a simple example of a bipartite system. These linear 
Langevin equations are derived from a discrete model for a two component cellular network estimating an external ligand concentration, which is the signal.
The two components of the network are receptors that can bind external ligands and internal proteins that play the role of memory \cite{meht12,gove14a,gove14,aqui14,bo15}.
This derivation starting with a physical model for a sensor allows us to provide a clear physical interpretation for the parameters showing up in the Langevin equations and
for the thermodynamic entropy production.   

The relation between sensory capacity and energy dissipation is also discussed. Particularly, as a main result we show  that 
if the sensory capacity is 1, the efficiency relating learning rate and rate of dissipation must be smaller than 1/2.
This result is valid for any bipartite process. The specific tradeoff between sensory capacity and efficiency for the 
coupled linear Langevin equations is analyzed in detail.

The paper in organized as follows. In Sec. \ref{sec:bipartite} we define discrete bipartite processes and the quantities calculated in the paper.  Sec. \ref{sec:two_component} contains
the derivation of the coupled linear Langevin equations from the microscopic model for a two component network. The analysis of the Langevin equations
is performed in Sec. \ref{sec:linear}. The general tradeoff between sensory capacity and efficiency is derived in Sec. \ref{sec:tradeoff}. We conclude in Sec. \ref{sec:conclusion}.
The continuum limit from a master equation to a Langevin equation in bipartite systems is presented in Appendix \ref{sec:A_master_to_langevin}. The uncertainty about the signal given 
the sensor state and the uncertainty given the sensor trajectory are calculated in Appendix \ref{sec:A_Un}.

\section{Bipartite Markov Processes and Sensory Capacity}
\label{sec:bipartite}
\subsection{Definition of bipartite systems}

A state of the signal is denoted by $x$ and a state of the sensor by $y$. We consider a quite 
general framework, where the basic assumptions are that the dynamics of the full system composed by the signal 
and the sensor is Markovian, the dynamics of 
the signal is not affected by the sensor whereas the dynamics of the sensor is affected by the signal, and
the signal alone is also Markovian. For a Markov jump process these assumptions imply the following 
transition rates from a state $(x,y)$ to a state $(x',y')$, 
\begin{equation}	
w_{yy'}^{xx'}\equiv
\begin{cases} 
 w^{xx'} & \quad \textrm{if $x\neq x'$ and $y= y'$}, \\
 w^{x}_{yy'} & \quad  \textrm{if $x= x'$ and $y\neq y'$},\\
 0 & \quad \textrm{if $x\neq x'$ and $y\neq y'$}. 
\end{cases}
\label{defrates}
\end{equation}
Such a Markov process, for which the two variables labeling a state cannot both change in a jump, is called bipartite \cite{bara13b}. The rates 
\eqref{defrates} correspond to a particular case of a bipartite process since $w^{xx'}$ is independent of $y$.
For bipartite systems in a steady state, which is the regime we consider in this paper, the stationary probability of state $(x,y)$ is written as $P(x,y)$. 
The marginals of this joint probability are defined as $P(x)\equiv \sum_{y}P(x,y)$ and $P(y)\equiv \sum_{x}P(x,y)$. The stationary conditional probabilities
read $P(x|y)\equiv P(x,y)/P(y)$ and $P(y|x)\equiv P(x,y)/P(x)$.

Key quantities in this paper are Shannon entropy and mutual information. The Shannon entropy associated with a random variable $A$ is
\begin{equation}
H[A]\equiv-\sum_{a}\mathcal{P}(A=a)\ln \mathcal{P}(A=a)
\label{shannonent}
\end{equation}
where $a$ is a specific realization of $A$ and $\mathcal{P}$ denotes a generic probability. 
The random variables $A$ can be the instantaneous state of the signal $x_t$ or of the sensor $y_t$. Furthermore, $A$ can be a full time series of the signal $\xtraj$ or of the sensor $\ytraj$.
In the first case, the sum in $a$ in Eq. \eqref{shannonent} is a sum over all possible states. In the second case, this sum corresponds to a functional integration over all possible trajectories.   
The conditional Shannon entropy of $A$ given another random variable $B$ is  
\begin{equation}
H[A|B]\equiv-\sum_{a,b}\mathcal{P}(A=a,B=b)\ln \mathcal{P}(A=a|B=b).
\label{shannoncond}
\end{equation}
The mutual information between $A$ and $B$ reads
 \begin{equation}
I[A{:}B]\equiv H[A]-H[A|B]= H[B]-H[B|A],
\label{mutualdef}
\end{equation}
where the second equality indicates that the mutual information is symmetric in the variables $A$ and $B$.
 
\subsection{Learning rate}

The learning rate is defined as \cite{bara14b}
\begin{equation}
l_\y\equiv\frac{H[x_t|y_t]-H[x_t|y_{t+\dd t}]}{\dd t},
\label{eq:lx_def}
\end{equation}
where here and in the following in all expressions that involve a $\dd t$ in the denominator the limit $\dd t\to 0$ is assumed. 
The learning rate quantifies the rate at which the sensor acquires information about the instantaneous signal state $x_t$, i.e., 
the rate at which the sensor reduces the uncertainty (as characterized by the conditional Shannon entropy) of the signal
due to its dynamics \cite{bara14b}. The learning rate can also be written in terms of mutual information 
\begin{equation}
l_\y=\frac{I[x_t{:}y_{t+\dd t}]-I[x_t{:}y_t]}{\dd t},
\label{eq:lx_def2}
\end{equation}
which is the rate at which the $y$ jumps increase the mutual information between the sensor $y$ and the signal $x$. This form
of the learning rate is also known as ``information flow'' \cite{alla09,horo14,horo14a}. 
Using the relations 
\begin{equation}
\begin{aligned}
 \mathcal{P}(\xtp=x'|x_t=x)&= w^{x x'}\dd t&&\text{for $x\neq x'$},\\
 \mathcal{P}(\ytp=y'|x_t=x,y_t=y)&= w^{x}_{y y'}\dd t&& \text{for $y\neq y'$}
\end{aligned}
\label{eq:def_rates}
\end{equation}
the learning rate \eqref{eq:lx_def} becomes
\begin{equation}
l_\y=\sum_{x,y,y'}P(x,y)w^{x}_{y y'}\ln \frac{P(x|y')}{P(x|y)}.
\label{eq:lx_master}
\end{equation}

In the steady state the learning rate is equal to the rate of Shannon entropy reduction of $x$ due to its coupling with $y$, which is defined as \cite{hart14}
\begin{equation}
h_\x\equiv\frac{H[\xtp|y_t]-H[x_t|y_t]}{\dd t}.
 \label{eq:hxdef}
\end{equation}
This conservation law comes from the relation  $\frac{\dd}{\dd t}H[x|y]\equiv h_\x-l_\y=0$ \cite{bara14b}, where $h_\x$ is the contribution 
due to the $x$ jumps, i.e.,  
\begin{equation}
h_\x=\sum_{x,x',y}P(x,y)w^{xx'}\ln \frac{P(x|y)}{P(x'|y)}.
\label{eq:hx_master}
\end{equation}
Since in the stationary state $H[y_{t+\dd t}]=H[y_{t}]$, the learning rate can also be written in the form
\begin{equation}
l_\y=h_\x=\frac{I[x_t{:}y_t]-I[\xtp{:}y_t]}{\dd t} 
\label{eq:hx_I}
\end{equation}
This expression is similar to the one used in \cite{stil12}, where within a discrete time formalism the term $I[\xtp{:}y_t]$ is identified as ``predictive power''.

\subsection{Sensory capacity and transfer entropy rate}

\begin{figure}%
 \centering%
 \includegraphics{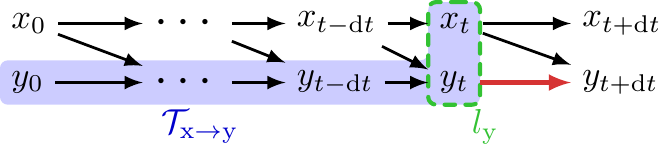}%
 \caption{(Color online) Learning rate versus transfer entropy rate. The learning rate takes into account only the instantaneous state $x_t$ (dashed green box) to infer the signal $x_t$, whereas the transfer 
 entropy $\mathcal{T}_{\x\to \y}$ takes into accout the trajectory highlighted by the blue shaded region.
 }%
 \label{fig:lx_vs_transfer_illustration}%
\end{figure}

Transfer entropy is an informational quantity that detects causal influence between two random variables \cite{schr00}. It plays an important role 
in the relation between information thermodynamics for causal networks \cite{ito13}, bipartite systems \cite{alla09,hart14,horo14a}, and feedback driven systems \cite{saga12a}.
The transfer entropy rate from the signal to the  sensor $\mathcal{T}_{\x\to \y}$  is defined as \cite{hart14}
\begin{align}
\mathcal{T}_{\x\to \y}
&\equiv \frac{H[\ytp|\ytraj]-H[\ytp|\ytraj,x_t]}{\dd t}\nonumber\\
&=\frac{I[\ytp{:}x_t|\ytraj]}{\dd t}\nonumber\\
&= \frac{H[x_t|\ytraj]-H[x_t|\ytp,\ytraj]}{\dd t}.
\label{eq:transfer_def}
\end{align}
In the third line the similarity with the learning rate \eqref{eq:lx_def} is explicit: the transfer entropy rate $\mathcal{T}_{\x\to \y}$ quantifies how much information 
the whole sensor trajectory $\ytraj$ contains about the instantaneous signal $x_t$, in contrast to the learning rate that considers only the instantaneous state $y_t$. 
This difference between the learning rate $l_\y$ and the transfer entropy rate $\mathcal{T}_{\x\to \y}$ is illustrated in Fig. \ref{fig:lx_vs_transfer_illustration}. 
The first line of Eq. \eqref{eq:transfer_def} contains the standard definition of transfer entropy from the signal to the sensor \cite{schr00}, which can be described as the reduction on the 
conditional Shannon entropy of $\ytp$ given $\ytraj$ by the further knowledge of the signal state $x_t$.

As shown in \cite{hart14} $l_\y\le \mathcal{T}_{\x\to \y}$, which simply means that the whole trajectory of the sensor $\ytraj$ contains more information about the 
instantaneous signal $x_t$ than the instantaneous state of the sensor $y_t$. Based on this inequality we propose the definition
\begin{equation}
C\equiv \frac{l_\y}{\mathcal{T}_{\x\to \y}}\le 1
\label{eq:eta_inf}
\end{equation}
that we call sensory capacity. If  $C=1$ the sensor has reached an information theoretical limit and its instantaneous state has the maximum possible information,
which is the information contained in the whole time series of the sensor. On a side note, as a result related to the fact that the full time series of a sensor contains 
more information about the signal than its instantaneous state, it has been shown that an information driven machine using the whole history 
of measurements can extract more work than a machine that only takes the last measurement into account \cite{baue14,bech15}. This increase in work extraction is 
characterized by a gain parameter that, like the sensory capacity, is positive and bounded by one.


\subsection{Thermodynamic entropy production and efficiency}

The thermodynamic entropy production \cite{seif12} for bipartite processes has two contributions. One is due to jumps that change the state of the signal, 
\begin{equation}
\sigma_\x \equiv\sum_{x,x'}P(x)w^{xx'}\ln\frac{w^{xx'}}{w^{x' x}}.
\label{eq:sigma_s_master}
\end{equation}
If the bare signal is an equilibrium process, which is the case for the examples considered in this paper, $\sigma_\x=0$. The second contribution
arises from jumps that change the state of the sensor, which reads
\begin{equation}
 \sigma_\y\equiv\sum_{x,y,y'}P(x,y)w^x_{y y'}\ln\frac{w^x_{yy'}}{w^x_{y' y}}.
\label{eq:sigma_x_master}
\end{equation}
The inequality $l_\y\le \sigma_\y$ leads to the efficiency \cite{bara14b}
\begin{equation}
\eta\equiv \frac{l_\y}{\sigma_\y}\le 1.
\label{eq:eta_th}
\end{equation}
This efficiency relates the rate at which the sensor learns about the signal with the rate of free energy dissipation, which is quantified by the thermodynamic entropy production.
For the model system in Sec. \ref{sec:two_component}, the entropy production has two terms. One is related to work done by the external signal and another to free energy dissipation
inside the cell.

\subsection{Upper bound on the transfer entropy, coarse-grained entropy production and coarse-grained learning rate}

We now recall the definition of further quantities that will be calculated in this paper. The first quantity is an upper bound on the transfer entropy rate  
\begin{equation}
\overline{\mathcal{T}}_{\x\to \y}\equiv\frac{H[y_{t+\dd t}|y_t]-H[y_{t+\dd t}|y_t,x_t]}{\dd t}
\ge \mathcal{T}_{\x\to \y}.
\label{eq:uppertransfer_def}
\end{equation}
An important property of this upper bound is that, unlike the transfer entropy rate,  it can be written in terms of the stationary distribution as \cite{hart14}
\begin{equation}
\overline{\mathcal{T}}_{\x\to \y}=\sum_{x,y,y'}P(x,y)w^x_{yy'}\ln\frac{w^x_{yy'}}{\overline{w}_{yy'}},
\label{eq:Tsxu_master}
\end{equation}
where 
\begin{equation}
\overline{w}_{yy'}\equiv \sum_{x}P(x|y)w^x_{yy'}.
\label{eq:eff_rates}
\end{equation}
The inequality  $\overline{\mathcal{T}}_{\x\to \y}\ge \mathcal{T}_{\x\to \y}$ is obtained by comparing Eq. \eqref{eq:transfer_def} with Eq. \eqref{eq:uppertransfer_def}, and 
using relations $H[y_{t+\dd t}|y_t]\ge H[\ytp|\ytraj]$ and $H[\ytp|\ytraj,x_t]=H[\ytp|y_t,x_t]$.

The coarse grained entropy production is obtained by integrating the variable $x$ out, leading to the expression \cite{espo12}
\begin{equation}
 \tilde\sigma_\y\equiv \sum_{yy'}P(y)\overline{w}_{yy'}\ln\frac{\overline{w}_{yy'}}{\overline{w}_{y' y}}\ge0.
 \label{eq:sigma_cg}
\end{equation}
This $\tilde\sigma_\y$ is a lower bound on the real entropy production, i.e., $\sigma_\y\ge\tilde\sigma_\y$ \cite{espo12}.


\subsection{Sensor with a memory}
\label{subsec:two_degrees_of_freedom}
\begin{figure}%
 \centering%
 \includegraphics{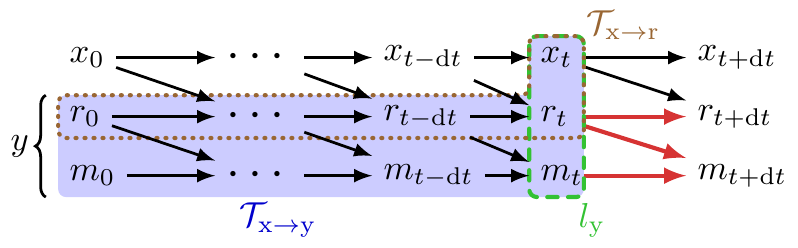}%
 \caption{(Color online) Illustration of the causal relation $x\to r\to m$ for a sensor $y=(r,m)$ composed of the first layer $r$ and the memory $m$.
 }
 \label{fig:lx_vs_transfer_2degrees}%
\end{figure}

We now consider a sensor with two degrees of freedom $y\equiv(r,m)$.
We assume that $r$ is the first degree of freedom directly sensing the signal $x$ and $m$ is a memory storing the information collected by $r$ (see \cite{bo15} for a similar setup).
The coarse-grained learning rate is defined as \cite{bara14b}
\begin{align}
 l_{\rr} & \equiv\frac{H[x_t|r_t]-H[x_t|r_{t+\dd t}]}{\dd t}\nonumber\\
 &  =\sum_{x,r,r',m}P(x,r,m)w^x_{(r,m)(r',m)}\ln\frac{P(x|r')}{P(x|r)},
\label{eq:coarsel} 
\end{align}
where $w^x_{(r,m)(r',m)}$ denotes the transition rate from $(x,r,m)$ to $(x,r',m)$. The rate at which $r$ alone learns about the signal $x$ is quantified by $l_{\rr}\le l_{\y}$ \cite{bara14b}. 
The transition rates 
then have the form
\begin{equation}	
w_{yy'}^{xx'}\equiv
\begin{cases} 
 w^{xx'} & \quad \textrm{if $x\neq x'$ and $y= y'$}, \\
 w^{x}_{r r'} & \quad  \textrm{if $x= x'$, $r\neq r'$  and $m=m'$},\\
 w_{(r,m)(r,m')} & \quad  \textrm{if $x= x'$, $r= r'$  and $m\neq m'$},\\
 0 & \quad \textrm{otherwise},
\end{cases}
\label{defrates2}
\end{equation}
where $y'= (r',m')$. The transitions rates \eqref{defrates2} imply the causal relation $x\to r\to m$, which is illustrated in Fig. \ref{fig:lx_vs_transfer_2degrees}.
Therefore, the coarse grained learning rate in Eq. \eqref{eq:coarsel} becomes
\begin{equation}
 l_{\rr}=\sum_{x,r,r'}P(x,r)w^x_{rr'}\ln\frac{P(x|r')}{P(x|r)}.
\label{eq:coarsel2} 
\end{equation}
Transition rates with three variables that do not change simultaneously in a jump, as in Eq. \eqref{defrates2}, 
form a tripartite system, which is a particular case of a multipartite Markov process \cite{horo15}. The transfer entropy in this case fulfills the relation  
\begin{equation}
\mathcal{T}_{\x\to \y}=\mathcal{T}_{\x\to \rr},
\label{transferyr}
\end{equation}
where
\begin{equation}
\mathcal{T}_{\x\to \rr} \equiv \frac{H[\rtp|\rtraj]-H[\rtp|\rtraj,x_t]}{{\dd t}}.
\label{eq:Txr_def}
\end{equation}
Relation \eqref{transferyr} means that the transfer entropy from the signal $x$ to the sensor $y=(r,m)$ is equal to the transfer entropy from $x$
to the first layer of the sensor $r$. This relation is a consequence of the causal relation $x\to r\to m$ and can be demonstrated as follows.

By defining $\bz_t\equiv(x_t,r_t,m_t)$ the conditional probability $\mathcal{P}(\bz_{t+\dd t}|\bz_t)$ can be written as  
\begin{equation}
 \mathcal{P}(\bz_{t+\dd t}|\bz_t)=\mathcal{P}(\xtp|x_t)\mathcal{P}(r_{t+\dd t}|x_t,r_t)\mathcal{P}(m_{t+\dd t}|r_t,m_t),
 \label{eq:P_factorize}
\end{equation}
which follows from the structure of the rates in Eq. \eqref{defrates2}. From the definition of the conditional Shannon entropy \eqref{shannoncond}, Eq. \eqref{eq:P_factorize} implies the following
relations
\begin{multline}
 H[\bz_{t+\dd t}|\bz_t]=\\
 H[\xtp|x_t]+H[r_{t+\dd t}|x_t,r_t]+H[m_{t+\dd t}|r_t,m_t],
\end{multline}
and
\begin{align}
H[\ytp|y_t,x_t]&\equiv
 H[\rtp,\mtp|r_t,m_t,x_t]\nonumber\\
 &=H[\rtp|r_t,x_t]+H[\mtp|r_t,m_t].
 \label{eq:trans_rm_last}
\end{align}
For large time $t$, the Markov property $\mathcal{P}(\bz_{t+\dd t}|\bz_t)=\mathcal{P}(\bz_{t+\dd t}|\{\bz_{t'}\}_{t'\le t})$ and \eqref{eq:P_factorize} lead to
\begin{align}
&H[\ytp|\ytraj]=
H[\rtp,\mtp|\rtraj,\mtraj]\nonumber\\
&\qquad=
H[\rtp|\rtraj]+H[\mtp|m_t,r_t].
\label{eq:trans_rm_first}
\end{align}
Finally, from Eqs. \eqref{eq:trans_rm_last} and \eqref{eq:trans_rm_first} we obtain the transfer entropy rate \eqref{eq:transfer_def} in the form 
\begin{align}
 \mathcal{T}_{\x\to\y}&=\frac{H[\ytp|\ytraj]-H[\ytp|y_t,x_t]}{\dd t}\nonumber\\
 &=\frac{H[\rtp|\rtraj]-H[\rtp|r_t,x_t]}{\dd t},
\end{align}
which after a comparison with \eqref{eq:Txr_def} yields the desired equality \eqref{transferyr}.

From the definition of the upper bound on the transfer entropy rate \eqref{eq:uppertransfer_def} and
Eq. \eqref{eq:P_factorize} we obtain
\begin{equation}
  \overline{\mathcal T}_{\x\to \y}= \frac{H[r_{t+\dd t}|r_t,m_t]- H[r_{t+\dd t}|r_t,x_t]}{\dd t}.
\end{equation}
Hence, the inequality $H[r_{t+\dd t}|r_t,m_t]\le H[r_{t+\dd t}|r_t]$ leads to
\begin{equation}
 \overline{\mathcal T}_{\x\to \y}\le \overline{\mathcal T}_{\x\to \rr},
\label{ineupp}
 \end{equation}
where
\begin{equation}
\overline{\mathcal{T}}_{\x\to \rr}\equiv\frac{H[r_{t+\dd t}|r_t]-H[r_{t+\dd t}|r_t,x_t]}{\dd t}.
\label{eq:uppertransfer_def2}
\end{equation}
Note that inequality \eqref{ineupp} is the opposite to what happens to the learning rate, i.e., $l_\rr\le l_\y$.
The chain of inequalities that summarizes the inequalities discussed in this section  involving learning rate, coarse grained learning rate, transfer entropy rates and upper bounds on
transfer entropy rates is given by
\begin{equation}
 l_\rr\le l_\y\le\mathcal{T}_{\x\to \rr}=\mathcal{T}_{\x\to \y}\le\overline{\mathcal T}_{\x\to \y}\le \overline{\mathcal T}_{\x\to \rr}.
 \label{eq:chain_ineq}
\end{equation}
The adaptation of the expressions from this section to the continuous limit, where the master equation becomes a Fokker-Planck equation, is presented in Appendix \ref{sec:A_master_to_langevin}. 

\section{Cellular two component network sensing an external ligand concentration}
\label{sec:two_component}

\begin{figure}
\centering
\includegraphics{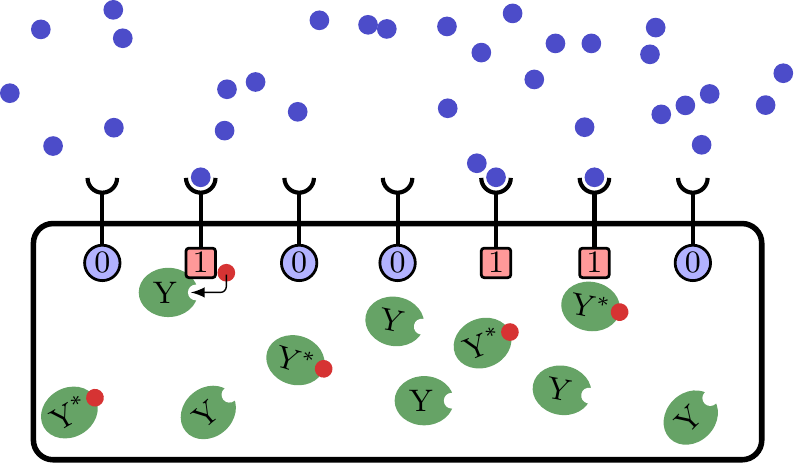}
 \caption{(Color online) Cellular two-competent network sensing an external ligand concentration.
 The total number of receptors is $N_{\rm b}=7$ and the number of occupied receptors is $n_{\rm b}=3$.
 The number of internal proteins, which constitute the memory, is $N_\y=10$ with $n_\y=4$ of them phosphorylated.
 The number of occupied receptors affects the transition rates related to the phosphorylation of internal proteins.}
 \label{fig:2component}
\end{figure}

As a physical realization of a sensor  we consider the cellular two component network sensing a fluctuating ligand concentration shown in Fig. \ref{fig:2component} (see \cite{bo15} for a similar setup). 
The signal $x$ is related to the external ligand concentration $s$ through the expression $x=\ln (s/s_0)$, where $s_0$ is some base concentration value.
The first layer of the two-component network, which is the degree of freedom directly sensing the external concentration,
is composed by the receptors. Each receptor can be either bound by a ligand or empty, with the possible values of the number of bound receptors given by $n_\rmb=0,1,\ldots,N_\rmb$, where $N_\rmb$ is the total number of receptors. 
The second layer of the two-component network is composed by internal proteins ${\rm Y}$ that can be phosphorylated to the state ${\rm Y}^*$. The number of proteins in this phosphorylated form takes the values $n_\y=0,1,\ldots,N_\y$, 
where $N_\y$ is the total number of proteins. This second degree of freedom is the memory of the sensor: the phosphorylation/dephosphorylation reaction rates depend on $n_\rmb$, whereas $n_\y$ has no
influence on the transition rates changing the number of occupied receptors. A state of the sensor is fully characterized by $y=(n_\rmb,n_\y)$.  

The rates with which the concentration changes are written as 
\begin{equation}
 w^{(1)}_{\pm}(x)=\frac{D_\x}{\dd x^2}\exp\left(\mp\frac{\omega_\x x}{2D_\x}\dd x\right),
 \label{eq:ws_pm}
\end{equation}
where $x$ is a multiple of $\dd x$ and the ``$+$'' sign indicates a jump from $x$ to $x+\dd x$ while 
the  ``$-$'' sign indicates a jump from $x$ to $x-\dd x$. 
As shown in Appendix \ref{sec:A_master_to_langevin}, the limit $\dd x\to 0$ yields the continuous Langevin equation 
\begin{equation}
\dot x_t=-\omega_\x x_t+\xi_t^\x,
\label{eq:langevin_signal_2component}
\end{equation}
for the dynamics of the signal. The white noise $\xi_t^\x$ fulfills the relation 
\begin{equation}
\avg{\xi_t^\x\xi_{t'}^\x}=2D_\x\delta(t-t'),
\label{noisedef}
\end{equation}
where the brackets denote an average over stochastic trajectories.

The number of occupied receptors changes with rates 
\begin{equation}
\begin{aligned}
 w^{(2)}_{+}(x,n_\rmb)&=
 \omega_{\rr}^+(x)[N_\rmb-n_\rmb]\\
 w^{(2)}_{-}(x,n_\rmb)&=
  \omega_{\rr}^-(x)n_\rmb,
\end{aligned}
\label{rates2}
\end{equation}
where $\omega_{\rr}^+(x)$ is the rate for the binding of a ligand to any free receptor and $\omega_{\rr}^-(x)$ is the rate for the unbinding of a ligand from any occupied receptor. These rates
fulfill the generalized detailed balance relation $\omega_{\rr}^+(x)/\omega_{\rr}^-(x)=\exp[\varDelta F(x)]$, where $\varDelta F(x)$ is the free energy difference between empty and occupied receptor
and $k_{\rm B}T\equiv1$ throughout.

The phosphorylation reaction of a single internal protein takes place with rates   
\begin{equation}
 {\rm Y}+{\rm ATP}\xrightleftharpoons[n_\rmb\kappa_-]{n_\rmb\kappa_+} {\rm Y}^*+{\rm ADP},
 \label{eq:phos}
\end{equation}
which are proportional to the number of bound receptors $n_\rmb$. Besides this chemical reaction the internal proteins can also be dephosphorylated through the reaction 
 \begin{equation}
 {\rm Y}^*\xrightleftharpoons[\nu_-]{\nu_+} {\rm Y}+{\rm P_i},
  \label{eq:dephos}
\end{equation}
where the rates are independent of $n_\rmb$. The rates in \eqref{eq:phos} and \eqref{eq:dephos} fulfill the relation $\ln[\kappa_+\nu_+/(\kappa_-\nu_-)]\equiv\varDelta\mu$, where 
$\varDelta\mu\equiv\mu_{\rm ATP}-\mu_{\rm ADP}-\mu_{\rm P_i}$ is the free energy liberated in one ATP hydrolysis. 
We define the total transition rates for individual proteins as
\begin{equation}%
\begin{aligned}%
 \omega^+_\m(n_\rmb)&\equiv n_\rmb\kappa_+  +\nu_-,\\
 \omega^-_\m(n_\rmb)&\equiv n_\rmb\kappa_-  +\nu_+.
\label{eq:hydrolysis_rates}%
\end{aligned}%
\end{equation}%
With these rates for the change of an individual protein we obtain the transition rates for a change in the variable $n_\y$, 
\begin{equation}
\begin{aligned}%
 w^{(3)}_{+}(n_\rmb,n_\y)&= \omega_{\m}^+(n_\rmb)[N_\y-n_\y],\\
 w^{(3)}_{-}(n_\rmb,n_\y)&=  \omega_{\m}^-(n_\rmb)n_\y.
\end{aligned}%
\label{rates3}
\end{equation}

The entropy production due to the sensor jumps $\sigma_\y$ has two contributions. The first is due to jumps 
that change the receptors occupancy
\begin{equation}
 \sigma_\rr=\sum_{\hidewidth x,n_\rmb\hidewidth}
 J_\rr(x,n_\rmb)\ln\frac{w^{(2)}_+(x,n_\rmb)}{w^{(2)}_-(x,n_\rmb+1)}
\label{entr}
 \end{equation}
where 
\begin{multline}
J_\rr(x,n_\rmb)\equiv\\ P(x,n_\rmb)w^{(2)}_+(x,n_\rmb)-P(x,n_\rmb+1)w^{(2)}_-(x,n_\rmb+1) 
\end{multline}
is the probability current.
The second is due to jumps that change the number of phosphorylated internal proteins 
\begin{equation}
  \sigma_\m=\sum_{n_\rmb,n_\y\hidewidth}J_\m(n_\rmb,n_\y)\ln\frac{w^{(3)}_+(n_\rmb,n_\y)}{w^{(3)}_-(n_\rmb,n_\y+1)}
 \label{entm}
 \end{equation}
where 
\begin{align}
J_\m(n_\rmb,n_\y)&\equiv P(n_\rmb,n_\y)w^{(3)}_+(n_\rmb,n_\y)\nonumber\\
&-P(n_\rmb,n_\y+1)w^{(3)}_+(n_\rmb,n_\y+1). 
\end{align}
The quantity $\sigma_\rr$ corresponds to the rate of dissipated heat due to binding and unbinding of ligands at different concentrations values. This 
dissipated heat is compensated by work that is done by the external signal. The quantity $\sigma_\y$ is the rate of dissipated 
free energy related to the consumption of ATP inside the cell. Actually, since we are not considering each individual link with the 
phosphorylation and dephosphorylation chemical reactions, but rather the total transition rates in Eq. \eqref{eq:hydrolysis_rates},   
$\sigma_\m$ is a lower bound on the rate of heat dissipated due to ATP consumption. A thorough discussion on the physical origin
of different terms in the entropy production for related models can be found in \cite{bara14b}.

As shown in Appendix \ref{sec:A_master_to_langevin}, taking the linear noise approximation and assuming a signal with small fluctuations, the transition rates in 
Eqs. \eqref{eq:ws_pm}, \eqref{rates2}, and \eqref{rates3} lead to the Langevin equations 
 \begin{align}%
  \dot{x}_t&= -\omega_\x x_t+\xi^\x_t&&\text{(signal)},\nonumber
 \\
 \dot{r}_t&= -\omega_\rr (r_t-x_t)+\xi^\rr_t&&\text{(sensor)}, \label{langevin3}
 \\
 \dot{m}_t&=-\omega_\m(m_t-r_t)+\xi_t^\m &&\text{(memory)},\nonumber
\end{align}%
where $\avg{\xi_t^i\xi_{t'}^j}=2D_i\delta_{ij}\delta(t-t')$ for $i,j= \x,\rr,\m$. The variable $r$ is related to the number of bound receptors, as shown in Eq. \eqref{eqa11}, and the memory $m$ to the number of phosphorylated
internal proteins, as shown in Eq. \eqref{eqa12}. The precise relations between the parameters in these equations and the transitions rates can be found in Appendix  \ref{sec:A_master_to_langevin}. There are three key points 
about these relations. First, for $\Delta \mu=0$, i.e., without free energy dissipation due to ATP hydrolysis 
inside the cell, the memory becomes decoupled from the receptor and has no information about the signal, which in Eq. \eqref{langevin3} implies $D_\m\to \infty$. Second, 
the noise amplitude $D_\rr$ is inversely proportional to the total number of receptors $N_\rmb$. Third, the noise amplitude $D_\m$ is inversely proportional to the total number of internal proteins $N_\y$.

\section{Sensory capacity and efficiency for model system}
\label{sec:linear}

\subsection{Bare sensor}
\label{sec:bare_sensor}

First we consider a bare sensor without memory, i.e., the Langevin equations \eqref{langevin3} without the variable $m$.
We use the subscript $\rr$ for the sensory capacity $C_\rr$ and the efficiency $\eta_\rr$ for the bare sensor of this subsection in order to 
differentiate it from the sensor with a memory analyzed in the next subsection. The corresponding Lyapunov equation for the covariance matrix 
\begin{equation}
 \bSigma=
 \begin{pmatrix}
  \varSigma_{\x\x}&\varSigma_{\x\rr}\\
  \varSigma_{\rr\x}&\varSigma_{\rr\rr}
 \end{pmatrix}
\equiv
\begin{pmatrix}
 \avg{x_t x_t}&\avg{x_t r_t}\\
  \avg{r_t x_t}&\avg{r_t r_t}
\end{pmatrix}
\end{equation}
reads \cite{gard04,bres14}
\begin{equation}
\dot\bSigma=-\mat{A}\bSigma-\bSigma\mat{A}^\T+2\mat{D},
 \label{eq:lypunov}
\end{equation}
where
\begin{equation}
 \mat{A}\equiv
\begin{pmatrix}
 \omega_\x&0\\
  -\omega_\rr&\omega_\rr
\end{pmatrix}\quad\text{and}\quad
\mat{D}\equiv
\begin{pmatrix}
 D_\x&0\\
 0&D_\rr
\end{pmatrix}.
\label{eq:2x2matrixAD}
\end{equation}
The steady state solution of \eqref{eq:lypunov} is
\begin{align}
\bSigma&=
\mathcal{E}_\x^2
\begin{pmatrix}
	 1
	&
	\frac{\nu_\rr}{\nu_\rr+1}
	\\
	\frac{\nu_\rr}{\nu_\rr+1}
	&
	\Big[\frac{\nu_\rr}{\nu_\rr+1}+\frac{B_\rr}{\nu_\rr}\Big]
\end{pmatrix},
\label{eq:covariance}
\end{align}
where $\mathcal{E}_\x^2\equiv D_\x/\omega_\x$ is the signal variance, $\nu_\rr\equiv\omega_\rr/\omega_\x$ and  $B_\rr\equiv D_\rr/D_\x$.
 \begin{figure}%
 \centering%
\includegraphics{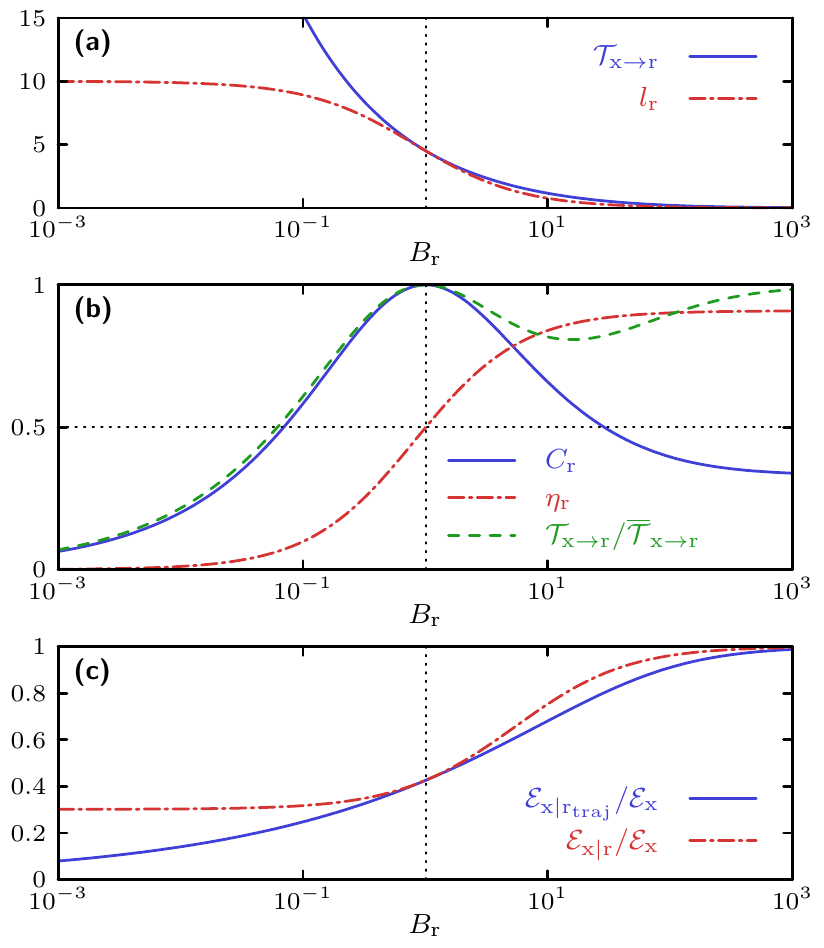}%
 \caption{(Color online) Sensor performance as function of sensor noise $B_\rr=D_\rr/D_\x$.
 (a) Transfer entropy $\mathcal{T}_{\x \to \rr}$ and learning rate $l_\rr$ are displayed.
 The vertical dotted line at $B_\rr=\nu_\rr^2/(\nu_\rr^2-1)$ indicates the value for which $C_\rr=1$, i.e., $l_\rr=\mathcal{T}_{\x \to \rr}$.
 (b) Efficiency ($\eta_\rr=l_\rr/\sigma_\rr$) and capacity ($C_\rr=l_\rr/\mathcal{T}_{\x \to \rr}$) of the bare sensor. At maximal capacity $C_\rr=1$ the efficiency is $\eta_\rr=1/2$ and $\mathcal{T}_{\x \to \rr}=\overline{\mathcal{T}}_{\x \to \rr}$. 
 (c) Comparison of errors. For $C_\rr=1$ the inequality $\mathcal{E}_{\x|\rr_{\rm traj}}\le \mathcal{E}_{\x|\rr}$ saturates.
 Parameters: $\omega_\x\equiv1,D_\x\equiv 0.1,\nu_\rr=\omega_\rr/\omega_\x\equiv10$.}%
 \label{fig:mathematica_error_vs_eta1}%
\end{figure}%

As shown in Appendix \ref{sec:A_master_to_langevin}, the learning rate is 
\begin{equation}
l_\rr=\omega_\x\left[\frac{\nu_\rr^3}{\nu_\rr^2+B_\rr(1+\nu_\rr)^2}\right]
 \label{eq:lx_linear}
\end{equation}
The transfer entropy rate for the linear Langevin equations \eqref{langevin3} is given by \cite{horo14a}
\begin{align}
\mathcal{T}_{\x\to \rr}&=\frac{\omega_\x}{2}\left(\sqrt{1+\frac{\nu_\rr^2}{B_\rr}}-1\right).
\label{eq:transfer_example}
\end{align}
The learning rate and transfer entropy rate as functions of $B_\rr$ are plotted in Fig. \ref{fig:mathematica_error_vs_eta1}(a).
Both quantities get smaller as the noise amplitude of the sensor gets larger. At an intermediate value of $B_\rr= \nu_\rr^2/(\nu_\rr^2-1)$ learning rate and transfer entropy become the same
leading to a sensory capacity $C_\rr=1$, as shown in Fig. \ref{fig:mathematica_error_vs_eta1}(b). 

Since the bare sensor does not have a memory there is no ATP consumption inside the cell and the entropy production is equal to the rate of work done by the external signal, which, as calculated in 
Appendix \ref{sec:A_master_to_langevin} in Eq. \eqref{eqappsigmar}, is
\begin{equation}
 \sigma_\rr=\omega_\x
 \frac{\nu_\rr^2}{B_\rr(1+\nu_\rr)}.
 \label{eq:sigma_x_linear}
\end{equation}
This entropy production decreases with $B_\rr$, i.e., a sensor with smaller noise amplitude, which can be obtained by increasing the number of receptors [see Eq. \eqref{noiseam2}], 
implies more energy dissipation.  In Fig. \ref{fig:mathematica_error_vs_eta1}(b) the thermodynamic efficiency is compared with sensory capacity. The efficiency increases with $B_\rr$. For 
$B_\rr= \nu_\rr^2/(\nu_\rr^2-1)$, where $C_\rr=1$, the efficiency is $\eta_\rr=1/2$. As we show in in Sec. \ref{sec:tradeoff} there is a general tradeoff between efficiency
and sensory capacity, with $C=1$ implying $\eta\le 1/2$.

The upper bound on the transfer entropy rate, calculated in Appendix \ref{sec:A_master_to_langevin},  reads 
\begin{equation}
 \overline{\mathcal{T}}_{\x\to \rr}=
\frac{\omega_\x\nu_\rr^2}{4B_\rr}\left[1-\frac{\nu_\rr^3}{\nu_\rr^3+\nu_\rr^2+B_\rr(1+\nu_\rr)^2}\right].
\label{eq:uppertransfer_example}
\end{equation}
This quantity has also been calculated in \cite{ito15}. Comparing the upper bound with the transfer entropy rate in Fig. \ref{fig:mathematica_error_vs_eta1}(b) we observe that for this model
when sensory capacity is one we have $l_\rr=\mathcal{T}_{\x\to \rr}=\overline{\mathcal{T}}_{\x\to \rr}$. This fact plays an important role
in the general tradeoff between sensory capacity and efficiency proved in Sec. \ref{sec:tradeoff}. 

In Appendix \ref{sec:A_Un} we define the uncertainties $\mathcal{E}_{\x|\rr}$ and $\mathcal{E}_{\x|\rr_\text{traj}}$ about the signal given the sensor state and the sensor trajectory, respectively.
As shown in Appendix \ref{sec:A_Un},  $\mathcal{E}_{\x|\rr_\text{traj}}^2$ is proportional to the transfer entropy rate  $\mathcal{T}_{\x\to \rr}$ and $\mathcal{E}_{\x|\rr}^2$ is proportional 
to the upper bound  $\overline{\mathcal{T}}_{\x\to \rr}$ for the present model. Hence, the equality between transfer entropy rate and upper bound for $C_\rr=1$ implies that both uncertainties are 
also the same, as shown in Fig. \ref{fig:mathematica_error_vs_eta1}(c).

\subsection{Memory increases sensory capacity}

For the regimes where the bare sensor does not reach a sensory capacity close to 1, it is possible to increase this sensory capacity by adding a memory to the bare sensor, which leads
to the third equation in \eqref{langevin3}.  The Lyapunov equation \eqref{eq:lypunov} for this case has the $3\times3$ matrices 
\begin{equation}
 \mat{A}=
 \begin{pmatrix}
  \omega_\x&0&0\\
  -\omega_\rr&\omega_\rr&0\\
  0&-\omega_\m&\omega_\m
 \end{pmatrix}
 \quad\text{and}\quad
 \mat{D}
 \equiv
 \begin{pmatrix}
 D_\x&0&0\\
 0&D_\rr&0\\
 0&0&D_\m
 \end{pmatrix}.
 \label{eq:3x3matrixAD}
\end{equation}
The stationary solution of \eqref{eq:lypunov} is too long to be displayed here.  

\begin{figure}%
 \centering%
 \includegraphics{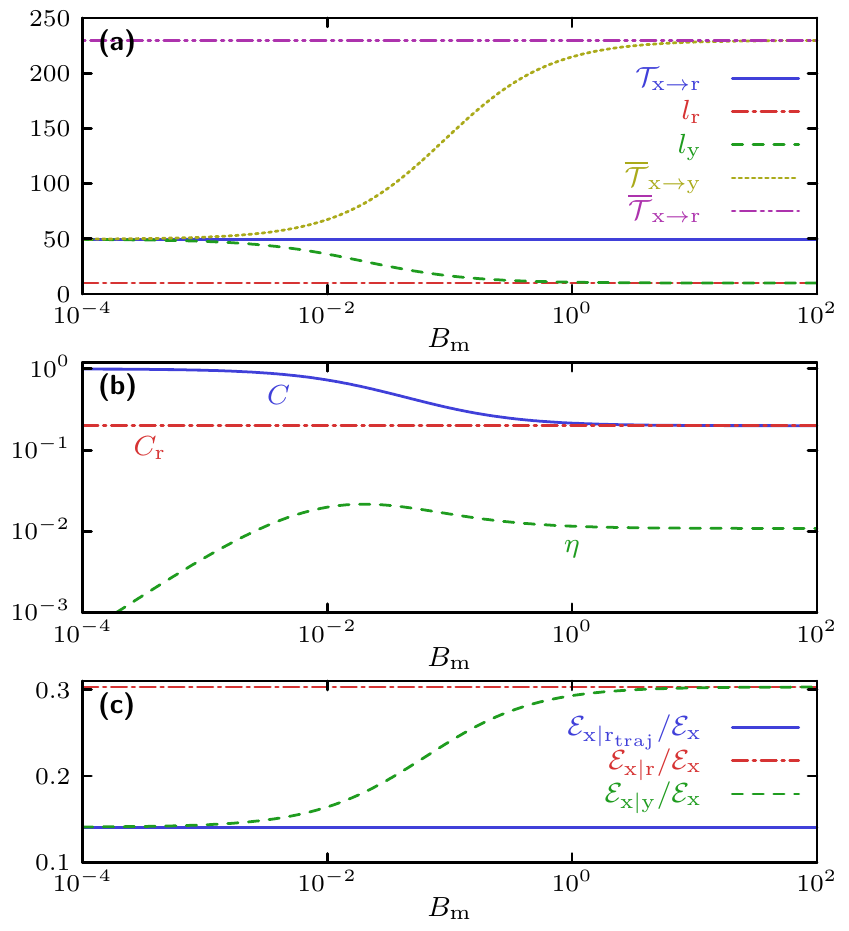}%
\caption{(Color online) Effect of a memory.
 (a) Transfer entropy $\mathcal{T}_{\x\to\rr}$, learning rate of the bare sensor $l_\rr$ and of the full sensor $l_\y$ (including the memory) as function of the memory noise $B_\m= D_\m/D_\x$. 
 The transfer entropy estimate $\overline{\mathcal{T}}_{\x\to\y}$ and the learning rate $l_\y$ approach $\mathcal{T}_{\x\to\rr}$ for $B_\m\to0$. 
 (b) Sensory capacities $C= l_{\y}/\mathcal{T}_{\x\to \rr}$ and $C_\rr= l_{\rr}/\mathcal{T}_{\x\to \rr}$
 in comparison with thermodynamical efficiency $\eta= l_{\y}/\sigma_{\y}$.
 (c) Effect of memory on error. The error  $\mathcal{E}_{\x|\y}$ corresponding to the full sensor state approaches the minimal error $\mathcal{E}_{\x|\rr_{\rm traj}}$ for $B_\m\to0$.
 Parameters: $\omega_\x\equiv1$, $D_\x\equiv10^{-1}$, $\nu_\rr=\omega_\rr/\omega_\x\equiv10$, $B_\rr=D_\rr/D_\x\equiv10^{-2}$, and $\nu_\m=\omega_\m/\omega_\x\equiv\sqrt{1+\nu_\rr^2/B_\rr}\simeq 100$.}%
 \label{fig:mathematica_error_vs_eta2}%
\end{figure}

The expression for the learning rate $l_\y$ is given in Appendix \ref{sec:A_master_to_langevin} in Eq. \eqref{lyexpression}.
As shown in Eq. \eqref{transferyr}, the addition of the memory does not change the transfer entropy $\mathcal{T}_{\x\to \y}=\mathcal{T}_{\x\to \rr}$ which remains as given by \eqref{eq:transfer_example}. 
The coarse grained learning rate $l_\rr$ is the learning rate for the bare sensor calculated in Eq. \eqref{eq:lx_linear}. The quantities $l_\y$, $l_\rr$ and $\mathcal{T}_{\x\to \rr}$  are plotted 
in Fig. \ref{fig:mathematica_error_vs_eta2}(a) as a function of the noise amplitude $B_\m\equiv D_\m/D_\x$.
For larger values of $B_\m$ the learning rate $l_\y$ becomes equal to $l_\rr$, the learning rate does now increase substantially with the addition of a memory with large noise amplitude.
By decreasing the noise amplitude  $l_\y$ increases until it reaches the transfer entropy $\mathcal{T}_{\x\to \rr}$ for small $B_\m$. 
Hence, the sensory capacity $C$ increases with decreasing  $B_\m$, as shown in Fig. \ref{fig:mathematica_error_vs_eta2}(b). 

The rate of free energy dissipation has now two contributions, i.e., $\sigma_\y=\sigma_\rr +\sigma_\m$. The $\sigma_\rr$ given by \eqref{eq:sigma_x_linear} corresponds to the work 
done by the external signal. The additional term, which is derived in Appendix \ref{sec:A_master_to_langevin} in Eq. \eqref{eq:A_sigma_m_linear}, is given by
\begin{equation}
 \sigma_\m=\omega_\x\frac{\nu_\m^2[\nu_\rr^2+B_\rr(1+\nu_\m)(1+\nu_\rr)]}{B_\m (1+\nu_\m)(1+\nu_\rr)(\nu_\m+\nu_\rr)},
 \label{eq:sigma_m_linear}
\end{equation}
where $\nu_\m\equiv \omega_\m/\omega_\x$.
This $\sigma_\m$ is a lower bound on the rate of dissipated free energy due to ATP consumption. From expression \eqref{eq:sigma_m_linear}, 
the decrease in the noise amplitude $D_\m$, which leads to an increase in sensory capacity, implies an increase in the rate of ATP consumption inside the cell.  
Adding a dissipative memory to a bare sensor can lead to an increase in sensory capacity. This increase corresponds to how much of the information 
about the trajectory $\rtraj$ is contained in the instantaneous state of the memory $m_t$.

\begin{figure}%
 \centering%
\includegraphics{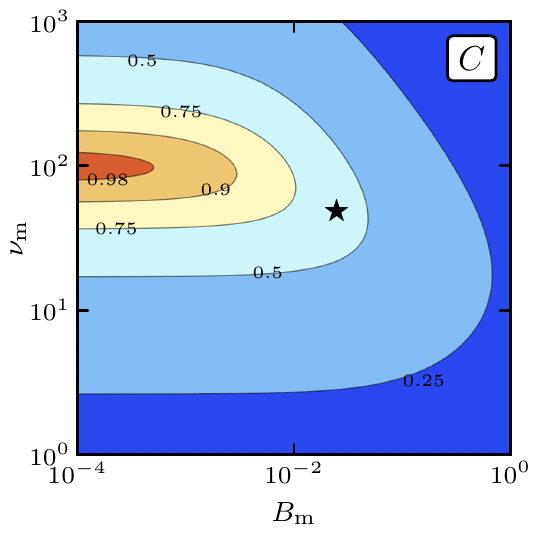}%
 \caption{(Color online)
 Effect of memory parameters $\nu_\m$ and $B_\m$ on the sensory capacity.
 For $\nu_\m=\sqrt{1+\nu_\rr^2/B_\rr}\simeq10^2$ and $B_\m\to0$ the capacity saturates ($C \to1$). The star ($\bigstar$) marks  the parameter $(\nu_\m^\bigstar,B_\m^\bigstar)$ for which the efficiency $\eta$ is maximal (here $\eta^\bigstar\simeq0.024$).
 The remaining parameter are chosen as in Fig. \ref{fig:mathematica_error_vs_eta2}.
 }
 \label{fig:mathematica_error_vs_eta3}
\end{figure}%

For fixed $B_\m$, the sensory capacity $C$ as a function of $\nu_\m\equiv\omega_\m/\omega_\x$ has a maximum, as shown in the contour plot in Fig. \ref{fig:mathematica_error_vs_eta3}. 
Therefore, for a given $\omega_\x$, which characterizes the time-scale of changes in the external signal, the memory has an optimal $\omega_\m$, which characterizes the time-scale of changes in the memory.
A sensory capacity close to 1 is reached for small $B_\m$ and  $\nu_\m\approx \sqrt{1+\nu_\rr^2/B_\rr}$, as indicated by the red region in Fig. \ref{fig:mathematica_error_vs_eta3}.

A larger $\sigma_\m$ leads to a lower efficiency, as shown in Fig. \ref{fig:mathematica_error_vs_eta2}(b). Adding a memory with a high rate of dissipation due to ATP
consumption can increase a low sensory capacity to the limit $C=1$. In this case when $C=1$ the efficiency is small due to the high dissipation of the memory. For example, 
the maximal efficiency that is achieved in the region plotted in Fig.  \ref{fig:mathematica_error_vs_eta3} is $\eta\simeq 0.024$. In this regime of high internal dissipation the efficiency 
does not seem to be a relevant quantity to characterize the performance of the sensor, which is rather given by sensory capacity. 

As shown in Appendix \ref{sec:A_Un}, for a sensor with a memory, the uncertainty taking the instantaneous state of the sensor into account is proportional to
the upper bound on the transfer entropy rate. As is the case of the transfer entropy, the uncertainty taking the full time series of the sensor into account 
does not change with the addition of the memory. Therefore, also for the present case $C=1$ implies that both uncertainties are equal, as shown in Fig \ref{fig:mathematica_error_vs_eta2}(c).

\section{Tradeoff between Sensory Capacity and Efficiency} 

\label{sec:tradeoff}

\subsection{Tradeoff for model system}

There are two situations for which the maximal sensory capacity $C=1$ can be reached. Either the parameters related to the signal and the first layer of the sensor are chosen in such a way that there is no
further information in the trajectory $\rtraj$ as compared to the instantaneous state $r_t$ or a dissipative memory is added to the sensor. In the first case, the efficiency 
is $\eta=1/2$ for $C=1$ and in the other case $\eta<1/2$ due to the extra dissipation inside the cell. 

\begin{figure}
 \includegraphics{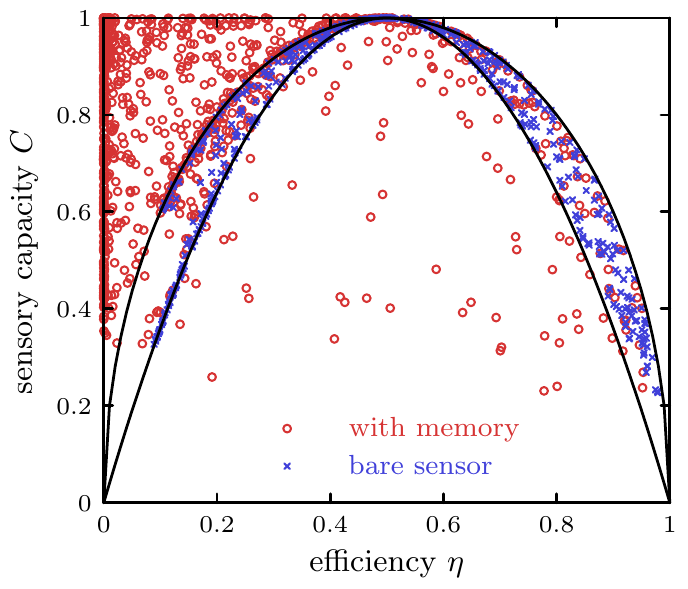}
 \caption{Trade-off between capacity $C$ and efficiency $\eta$. The parameters for the bare sensor $\nu_\rr$ and $B_\rr$ are chosen at random with $10^{-1}\le\nu_\rr,B_\rr\le10^2$. 
 For the sensor with memory, in addition, the parameters $\nu_\m$ and $B_\m$ are chosen in the same way. The solid lines indicate the bounds $4\eta(1-\eta)\le C\le 2\sqrt{\eta(1-\eta)}$ 
 for bare sensor. Our numerics indicates that the upper bound $C\le 2\sqrt{\eta(1-\eta)}$ is also valid for the sensor with memory for $\eta\ge 1/2$.
}
 \label{fig:etaeta}
\end{figure}

The tradeoff between sensory capacity and efficiency for the model system in Eq. \eqref{langevin3} is shown in Fig. \ref{fig:etaeta}. 
For the bare sensor we obtain the bounds 
\begin{equation}
4\eta_\rr(1-\eta_\rr)\le C_\rr\le 2\sqrt{\eta_\rr(1-\eta_\rr)},
\label{boundCeta}
\end{equation}
which are derived in the following way. From \eqref{eq:lx_linear} and \eqref{eq:sigma_x_linear} the efficiency reads
\begin{equation}
 \eta_\rr=\frac{l_\rr}{\sigma_\rr}=\frac{B_\rr\nu_\rr(1+\nu_\rr)}{\nu_\rr^2+B_\rr(1+\nu_\rr)^2},
 \label{eq:eta_bare_sensor}
\end{equation}
and from \eqref{eq:lx_linear} and \eqref{eq:transfer_example} the sensory capacity reads
\begin{equation}
 C_\rr=\frac{l_\rr}{\mathcal{T}_{\x\to\rr}}=\frac{2\nu_\rr^3}{[\nu_\rr^2+B_\rr(1+\nu_\rr^2)][\sqrt{1+\nu_\rr^2/B_\rr}-1]}.
  \label{eq:capacity_bare_sensor}
\end{equation}
The upper (lower) bound in Eq. \eqref{boundCeta} is obtained by maximizing (minimizing) the capacity \eqref{eq:capacity_bare_sensor} with respect to the variables $\nu_\rr,B_\rr\ge0$ with the constraint 
that \eqref{eq:eta_bare_sensor} is fixed. Most prominently, the scatter plot in Fig. \ref{fig:etaeta} shows that the upper bound in Eq. \eqref{boundCeta} also applies for the full sensor with 
a memory in the region $\eta\ge 1/2$.

\subsection{General proof}

We now prove as a general trade-off between sensory capacity and efficiency: a sensory capacity $C=1$ implies $\eta\le 1/2$. Our 
proof depends on the reasonable assumption that for any sensor it is possible to create a fictitious memory such that the instantaneous state of
the fictitious sensor, composed of the sensor and the fictitious memory, contains the whole history of the sensor. From the calculations for the model system
in Sec. \ref{sec:linear}, we expect this fictitious memory to have two general characteristics. First, it must be precise. For the model system this precision is characterize by a small $D_\m$ in Eq. \eqref{langevin3},
which can be achieved for the case the total number of proteins inside the cell is very large, i.e., the memory has a large number of possible states. Second, the time scale for changes in states of the fictitious memory must be
tuned to some optimal value. For the model system this time scale is characterized by $\omega_\m$ in Eq. \eqref{langevin3}. For a system that is more elaborate than our model system one can think of a multicomponent memory
with the time-scale of each component optimally tuned to store information about a certain part of the sensor. 

From the chain of inequalities, summarized in \eqref{eq:chain_ineq}, adding the memory raises the learning rate and lowers the upper bound on transfer 
entropy rate. In a first step, we impose that (i) $C=1$ and (ii) that the transfer entropy rate is equal to the upper bound, i.e.,  $l_\y=\mathcal{T}_{\x\to\y}=\overline{\mathcal T}_{\x\to \y}$.
From relations \eqref{eq:lx_master} and \eqref{eq:Tsxu_master} we obtain
\begin{multline}
 \overline{\mathcal{T}}_{\x\to\y}-l_\y\\=\sum_{y,y'}P(y)\sum_xP(x|y)w^x_{yy'}\ln\frac{P(x|y)w^x_{yy'}}{P(x|y')\overline{w}_{yy'}}\ge0,
\end{multline}
where the log sum inequality above is saturated if and only if the term inside the logarithm is independent of  $x$ \cite{cove06}.
Hence, if $\overline{\mathcal{T}}_{\x\to\y}=l_\y$ then the rates obey
\begin{equation}
 w^x_{yy'}=\frac{P(x|y')}{P(x|y)}\overline{w}_{yy'}.
\label{eq:rates_tradeoff}
\end{equation}
With this restriction, Eq. \eqref{eq:lx_master} and Eq. \eqref{eq:sigma_cg}, the entropy production \eqref{eq:sigma_x_master} becomes
\begin{equation}
\sigma_\y=2l_\y+\tilde{\sigma}_\y.
\label{eq:tradeoff_pre}
\end{equation}
The efficiency \eqref{eq:eta_th} then reads 
\begin{equation}
\eta_\y=\frac{l_\y}{\sigma_\y}=\frac{1}{2}\frac{\sigma_\y-\tilde{\sigma}_\y}{\sigma_\y}\le \frac{1}{2},
\label{eq:tradeoff}
\end{equation}
where we used $\sigma_\y\ge\tilde{\sigma}_\y$. Hence, if $C=1$ and $\mathcal{T}_{\x\to\y}=\overline{\mathcal T}_{\x\to \y}$, the efficiency fulfills $\eta_\y \le  1/2$.

We now demonstrate that $C=1$ indeed implies $\mathcal{T}_{\x\to\y}=\overline{\mathcal T}_{\x\to \y}$, which completes the proof of 
the tradeoff. A fictitious memory $\alpha$ is added to the sensor $y$.
The transitions rates are now of the form of Eq. \eqref{defrates2} with $y$ replacing $r$ and $\alpha$
replacing $m$. The learning rate of this fictitious sensor composed of $z=(y,\alpha)$ reads   
\begin{equation}
l_{\rm z}=\sum_{x,x',y,\alpha}P(x,y,\alpha)w^{xx'}\ln\frac{P(x,y,\alpha)}{P(x',y,\alpha)},
\label{eq:ly_tradeoff}
\end{equation}
where we used Eqs. \eqref{eq:hx_master} and \eqref{eq:hx_I}.
Within this fictitious sensor $l_\y$ is a coarse-grained learning rate and the difference between $l_{\rm z}$ and $l_\y$ reads 
\begin{equation}
l_{\rm z}-l_\y= \sum_{x,x',y}P(x,y)w^{xx'}\sum_\alpha P(\alpha|x,y)\ln\frac{P(\alpha|x,y)}{P(\alpha|x',y)}\ge 0.
\label{eq:tradeoff_ldiff}
\end{equation}
The assumption $C=1$ implies $l_\y=l_{\rm z}$. The above inequality is saturated if and only if
$P(\alpha|x,y)=P(\alpha|x',y)=P(\alpha|y)$, yielding $P(x|y,\alpha)=\frac{P(y)P(x|y)P(\alpha|x,y)}{P(y)P(\alpha|y)}=P(x|y)$. This relation leads to
\begin{equation}
H[x_t|y_t,\alpha_t]=H[x_t|y_t].
\label{eq:Hcond_tradeoff_pre}
\end{equation}
The fictitious memory $\alpha$ is unspecified and the key assumption for our demonstration is that it is always possible for any sensor $y$ to 
find a fictitious memory $\alpha$ that fulfills the relation 
\begin{equation}
H[x_t|y_t,\alpha_t]=H[x_t|\ytraj]. 
\label{keyass}
\end{equation}
If we choose such fictitious memory then equality \eqref{eq:Hcond_tradeoff_pre} leads to 
\begin{equation}
H[x_t|y_t]=H[x_t|\ytraj].
\label{eq:Hcond_tradeoff}
\end{equation}
Hence, if it is possible to find a fictitious memory that fulfills \eqref{keyass}, then  $C=1$ implies \eqref{eq:Hcond_tradeoff}.
From \eqref{eq:Hcond_tradeoff} we obtain $I[x_t{:}\ytraj]=I[x_t{:}y_t,y_{t-\dd t}]=I[x_t{:}y_t]$.
The learning rate in the form \eqref{eq:hx_I} can be rewritten as
\begin{align}
 l_\y&=\frac{I[x_t{:}y_t]-I[x_{t+\dd t}{:}y_t]}{\dd t }\nonumber\\
 &=\frac{I[x_{t+\dd t}{:}\ytp]-I[x_{t+\dd t}{:}y_t]}{\dd t}\nonumber\\
 &=\frac{I[x_{t+\dd t}{:}\ytp,y_t]-I[x_{t+\dd t}{:}y_t]}{\dd t},
\label{eq:ly_C=1}
\end{align}
where we used the steady state property $I[x_{t+\dd t}{:}\ytp]=I[x_{t}{:}y_t]$ from the first to the second line. 
Inserting the conditional probabilities in terms of rates from Eq. \eqref{eq:def_rates} into Eq. \eqref{eq:ly_C=1},
leads to the completion of the proof, i.e., 
\begin{align}
 l_\y&=\sum_{x,y,y'}P(x,y)w^x_{yy'}\ln\frac{w^x_{yy'}}{\overline{w}_{yy'}}=\overline{\mathcal T}_{\x\to\y}.
\label{eq:ly_Txyu_C=1}
\end{align}
Summarizing, we have demonstrated that $C=1\Rightarrow H[x_t|y_t]=H[x_t|\ytraj]\Rightarrow l_\y=\overline{\mathcal T}_{\x\to\y}\Rightarrow C=1$.
This proof also implies that whenever $C=1$ then the upper bound is also equal to the transfer entropy rate, i.e., $l_\y=\mathcal{T}_{\x\to\y}=\overline{\mathcal T}_{\x\to \y}$.
For the coupled linear Langevin equations analyzed in Sec. \ref{sec:linear} this equality between transfer entropy rate and its upper bound implies the equality between the uncertainty about the external signal
that are estimated with the instantaneous state of the sensor and the uncertainty  that is estimated with the full time series of the sensor, as shown in Appendix \ref{sec:A_Un}. For general systems, it remains to be seen 
whether $C=1$ implies that both uncertainties are the same.

\section{Conclusion}
\label{sec:conclusion}

We have introduced the quantity sensory capacity, which provides a measure for the performance of a sensor that follows an external signal. Specifically, the maximal sensory 
capacity $C=1$ means that the instantaneous state of the sensor contains the same amount of information about the signal as the full time-series of the sensor. As we have shown with 
the coupled linear Langevin equations in Sec. \ref{sec:linear} a high sensory capacity can be achieved in two ways. First, for a bare sensor without 
a memory layer the parameters related to the sensor can be tuned in such a way that $C=1$. In this case there is no further information available in
the full time series of the degree of freedom directly sensing the signal. Second, the more interesting case is when the full time series of this first degree of freedom
has more information than its instantaneous state. By adding a memory, which is a second degree of freedom that is influenced by the first degree of freedom but does not react back on it,
the sensory capacity can be raised to $C=1$. This increase in sensory capacity quantifies how much information about the time-series of the sensor
is stored in the instantaneous state of the memory.

The coupled linear Langevin equations have been derived from a cellular two component network sensing an external ligand concentration, which is the signal. Within this physical 
realization of a sensor the first layer of the sensor are the receptors that bind external ligand and the memory is composed of internal proteins that can be phosphorylated. 
We have shown that the thermodynamic entropy production quantifying dissipation has two terms: work done by the external process due to binding and unbinding at different concentrations
and dissipation inside the cell due to ATP hydrolysis. Adding a memory that increases the sensory capacity of a sensor from a low value to a value close to
one requires a high rate of dissipation inside the cell. Sensory capacity is particularly interesting in this regime of high dissipation, where the efficiency is very low 
and, therefore, does not characterize well the performance of the sensor.

Finally, we have demonstrated a general tradeoff between sensory capacity and efficiency. A sensory capacity $C=1$ implies an efficiency $\eta\le 1/2$. The limit
$\eta=1/2$ is achieved for a bare sensor with its parameters optimally tuned so that $C=1$. If these parameters are not optimally tuned, 
$C=1$ is possible only with an additional memory that leads to extra dissipation in relation to the bare sensor, which implies $\eta<1/2$.

This tradeoff relation between the two bounded dimensionless quantities $C$ and $\eta$ provides a further link between information theory and thermodynamics.
The sensory capacity $C$ as a ratio between learning rate and transfer entropy rate is of purely information theoretic origin whereas the efficiency $\eta$ as 
a ratio between learning rate and entropy production contains input from both fields. As a perspective for future work, the role of nonlinearities in
these figures of merit could be explored in more complex models. 

An experimental realization verifying the second law for a sensor that involves the rate of dissipated heat and the learning rate is still lacking.
A good candidate for such an experiment is a colloidal particle, which is the sensor, subjected to an external potential that is varied stochastically.
An experiment with a sensor that has an internal memory seems to be even more challenging.

\appendix

\section{From Master Equation to Langevin Equation in bipartite processes} 

\label{sec:A_master_to_langevin}
\subsection{Linear noise approximation}

We consider a vector $\bz=(z_1,\ldots,z_d)$ determining the state of the system. Comparing with Sec. \ref{sec:bipartite}, the first component is related to the signal, i.e., $z_1=x$.
The other components are related to the sensor. If the sensor has only one component $r$ then $z_2=r$. A sensor with a memory also has a second component $y=(r,m)$, leading
to $z_3=m$. For the variable $z_1=x$ we denote the transition rate $w^{xx'}=\omega_\pm^{(1)}(\bz)$ for $x'=x\pm\dd x$, where $\dd x$ corresponds to an infinitesimal change
in the variable $x$. The master equation is written as 
\begin{align}
 \dot P(\bz)&=\sum_{i=1}^d\Big[w_+^{(i)}(\bz -\dd \bz_i)P(\bz -\dd \bz_i)-w_+^{(i)}(\bz)P(\bz)\Big]\nonumber\\
&+\sum_{i=1}^d\Big[w_-^{(i)}(\bz +\dd \bz_i)P(\bz +\dd \bz_i)-w_-^{(i)}(\bz)P(\bz)\Big].
\label{eq:A_master_grid}
\end{align}
With the  approximation
\begin{multline}
w_\pm^{(i)}(\bz \mp\dd \bz_i)P(\bz \mp\dd \bz_i)\simeq
w_\pm^{(i)}(\bz)P(\bz)\\
\mp \dd z_i\frac{\del }{\del z_i}w_\pm^{(i)}(\bz)P(\bz)+\frac12\dd z_i^2\frac{\del^2}{\del z_i^2}w_\pm^{(i)}(\bz)P(\bz),
 \label{eq:FP_LNA_pre}
\end{multline}
the master equation \eqref{eq:A_master_grid} turns into the Fokker Planck equation
\begin{equation}
\dot\rho(\bz)=-\sum_i\frac{\del}{\del z_i}J_i(\bz),
 \label{eqfp}
\end{equation}
where in the continuous limit $P(\bz)\to \rho(\bz)\prod_i\dd z_i$. The probability current reads
\begin{equation}
J_i(\bz)\equiv D_i(\bz)F_i(\bz)\rho(\bz)-\frac{\del}{\del z_i}D_i(\bz)\rho(\bz),
\label{eq:A_current_def}
\end{equation}
where
\begin{equation}
D_i(\bz)F_i(\bz)\equiv\dd z_i\Big[w_+^{(i)}(\bz)-w_-^{(i)}(\bz)\Big],
\label{eq:A_gen_drift}
\end{equation}
and
\begin{equation}
D_i(\bz)\equiv\frac{\dd z_i^2}{2}\Big[w_+^{(i)}(\bz)+w_-^{(i)}(\bz)\Big]
\label{eq:A_gen_diff}
\end{equation}
Within the Ito interpretation \cite{gard04,bres14}, the Fokker-Planck equation \eqref{eqfp} corresponds to the Langevin  equation
\begin{equation}
 \dot z_{i,t}=D_i(\bz_t)F_i(\bz_t)+\xi_t^i,
 \label{langevingen}
\end{equation}
where $\avg{\xi_t^i\xi_{t'}^j}=2D_i(\bz)\delta_{ij}\delta(t-t')$. The $\delta_{ij}$ term in this last equation is a direct consequence of the bipartite (or multipartite) structure 
of the transition rates.

\subsection{Two component network with a weakly fluctuating signal }

The linear noise approximation  for the specific model of Sec. \ref{sec:two_component} is valid  
in  the limit $N_\y,N_\rmb\gg1$ and $\dd x\to 0 $. In this case, from the transition rates \eqref{eq:ws_pm}, \eqref{rates2}, and \eqref{rates3}, 
the Langevin equation \eqref{langevingen} becomes
\begin{align}
 \dot{x}_t&=-\omega_{\x}x_t+\xi^\x_t,\nonumber\\
 \dot{n}_{\rmb}(t)&=\omega_{\rr}^+(x_t)N_\rmb-\left[\omega_{\rr}^+(x_t)+\omega_{\rr}^-(x_t)\right]n_\rmb(t)+\xi^\rmb_t,\nonumber\\
 \dot{n}_{\y}(t)&=\omega_{\y}^+(n_\rmb(t))N_\y-\left[\omega_{\m}^+(n_\rmb(t))+\omega_{\m}^-(n_\rmb(t))\right]n_\y(t)\nonumber\\
 &\quad+\xi^\y_t.
 \label{langevin1}
\end{align}
From Eq. \eqref{eq:A_gen_diff},  the noise terms $\xi^\rmb_t$ and $\xi^\y_t$ fulfill a relation similar to \eqref{noisedef}, with amplitudes   
\begin{equation}
 \begin{aligned}
D_\rmb(x,n_\rmb)&=\frac{1}{2}\Big[\omega_{\rr}^+(x)(N_\rmb-n_\rmb)+\omega_{\rr}^-(x)n_\rmb\Big],\\
D_\y(n_\rmb,n_\y)&=\frac{1}{2}\Big[\omega_{\m}^+(n_\rmb)(N_\y-n_\y)+\omega_{\m}^-(n_\rmb)n_\y\Big],
 \end{aligned}%
 \label{noiseam}%
\end{equation}%
respectively. 

If the fluctuations of the signal are small such that $x$ stays close to the value $x=0$ we can apply 
the following expansion
\begin{equation}
 N_\rmb\omega_{\rr}^+(x)/[\omega_{\rr}^+(x)+\omega_{\rr}^-(x)]\equiv n_\rmb^*+\alpha_1x+\mathrm{O}(x)^2
 \label{eq:seires1}
\end{equation}
where $n_\rmb^*\equiv N_\rmb\omega_{\rr}^+(0)/[\omega_{\rr}^+(0)+\omega_{\rr}^-(0)]$ and $\alpha_1$ is the first derivative 
evaluated at $x=0$. For $n_\rmb-n_\rmb^*$ small, 
\begin{align}
N_\y\omega_{\m}^+(n_\rmb)/[\omega_{\m}^+(n_\rmb)+\omega_{\m}^-(n_\rmb)]&\equiv n_\y^*+\alpha_2(n_\rmb-n_\rmb^*)\nonumber\\
&\quad+\mathrm{O}(n_\rmb-n_\rmb^*)^2
\label{eq:seires2}
\end{align}
where $n_\y^*\equiv N_\y\omega_{\m}^+(n_\rmb^*)/[\omega_{\m}^+(n_\rmb^*)+\omega_{\m}^-(n_\rmb^*)]$ and $\alpha_2$ is the first derivative 
evaluated at $n_\rmb=n_\rmb^*$. In the limit where Eqs. \eqref{eq:seires1} and \eqref{eq:seires2} are valid, the Langevin equations \eqref{langevin1}
become 
\begin{align}
 \dot{x}_t&=-\omega_{\x}x_t+\xi^\x_t\nonumber\\
 \dot{n}_{\rmb}(t)&= \omega_\rr\Big[n_\rmb^*+\alpha_1 x_t-n_\rmb(t)\Big]+\xi^\rmb_t \label{eq_ny_langevin_lna_weak}\\
 \dot{n}_{\y}(t)&=\omega_\m\Big[n_\y^*+\alpha_2(n_\rmb(t)-n_\rmb^*)-n_\y(t)\Big]+\xi^\y_t,
\nonumber
\end{align}
where
\begin{equation}
\begin{aligned}
 \omega_\rr&\equiv\omega_{\rr}^+(0)+\omega_{\rr}^-(0)\\
 \omega_\m&\equiv\omega_{\m}^+(n_\rmb^*)+\omega_{\m}^-(n_\rmb^*).
\end{aligned}
\end{equation}
Furthermore, the noise amplitudes in Eq. \eqref{noiseam} become
\begin{equation}
\begin{aligned}
D^*_\rmb\equiv D_\rmb(0,n_\rmb^*)&=\frac{\omega_\rr}{N_\rmb}n_\rmb^*(N_\rmb-n_\rmb^*),\\
D^*_\y\equiv D_\y(n_\rmb^*,n_\y^*)&=\frac{\omega_\m}{N_\y}n_\y^*(N_\y-n_\y^*).
\label{noiseam2}
\end{aligned}
\end{equation}
The explicit form of the parameter
$\alpha_1$ in \eqref{eq:seires1} is
\begin{equation}
\alpha_1=\frac{n_\rmb^*(N_\rmb-n_\rmb^*)}{N_\rmb}\frac{\del \varDelta F(x)}{\del x}
\end{equation}
and $\alpha_2$ in Eq. \eqref{eq:seires2} is
\begin{equation}
\alpha_2=\frac{n_\y^*(N_\y-n_\y^*)}{N_\y}\left[\frac{\kappa_+\nu_+-\kappa_-\nu_-}{(n_\rmb^*\kappa_++\nu_-)(n_\rmb^*\kappa_-+\nu_+)}\right],
\end{equation}
as obtained from \eqref{eq:hydrolysis_rates}. Hence, for $\varDelta\mu=\ln[\kappa_+\nu_+/(\kappa_-\nu_-)] =0$ this last parameter is  $\alpha_2=0$, i.e.,
the memory level in Eq. \eqref{eq_ny_langevin_lna_weak} is not affected by the number of occupied receptors. Therefore, ATP consumpation
is necessary in order for the memory to be able to store information about the signal.    

The linear Langevin equations can be further simplified with the transformations  
\begin{equation}
 r_t\equiv\frac{n_\rmb(t)-n_\rmb^*}{\alpha_1}
 \label{eqa11}
\end{equation}
and 
\begin{equation}
 m_t\equiv\frac{n_\y(t)-n_\y^*}{\alpha_1\alpha_2}.
\label{eqa12}
 \end{equation}
With these variables the Langevin equations \eqref{eq_ny_langevin_lna_weak} become Eq. \eqref{langevin3},
with the noise amplitudes \eqref{noiseam2} transformed to
\begin{equation}
\begin{aligned}
 D_\rr&=D^*_\y/\alpha_1^2,\\
 D_\m&=D_\y^*/(\alpha_1\alpha_2)^2.
\end{aligned}
\end{equation}

\subsection{Quantities in the continuum limit}

We consider a vector $(z_1,z_2,z_3)=(x,r,m)$ with transition rates
\begin{align}
\omega^{(1)}_\pm(\bz)&\equiv\frac{D_\x}{\dd x^2}\exp\left[\pm\frac{ F_\x(x)\dd x}{2}\right],
\label{eq:A_srates}\\
\omega^{(2)}_\pm(\bz)&\equiv\frac{D_\rr}{\dd r^2}\exp\left[\pm\frac{ F_\rr(x,r)\dd r}{2}\right],
\label{eq:A_rrates}\\
\omega^{(3)}_\pm(\bz)&\equiv\frac{D_\m}{\dd m^2}\exp\left[\pm\frac{ F_\m(r,m)\dd m}{2}\right],
\label{eq:A_mrates}
\end{align}
where the the diffusion constants $D_i$ are assumed to be independent of $(x,r,m)$.  The following relations are obtained by taking their
expressions for the discrete case in Sec. \ref{sec:bipartite} and then taking the continuous limit $(\dd x,\dd r,\dd m)\to0$, where the probability is replaced by a density, i.e.,
$P(x,r,m)\to\rho(x,r,m)\dd x \dd r\dd m$.

\textit{Learning rate} --
From Eqs. \eqref{eq:FP_LNA_pre} and \eqref{eq:A_current_def} the learning rate \eqref{eq:lx_master} becomes
\begin{align}
 l_\y&=\int\dd x\int\dd r\int\dd mJ_\rr(x,r,m)\frac{\del}{\del r}\ln\rho(x|r,m)\nonumber\\
 &+\int\dd x\int\dd r\int\dd mJ_\m(x,r,m)\frac{\del}{\del m}\ln\rho(x|r,m),
 \label{eq:A_lx_langevin}
\end{align}
where $\rho(x|r,m)\equiv\rho(x,r,m)/[\int\rho(\tilde x,r,m)\dd \tilde x]$. This expression can also be
found in \cite{horo15}, where the learning rate is called information flow. Integration by parts
and the steady state property $\del_x J_\x+\del_r J_\rr+\del_m J_\m=0$ leads to the alternative expression
\begin{equation}
 l_\y=-\int\dd x\int\dd r\int\dd m J_\x(s,r,m)\frac{\del}{\del x}\ln\rho(x|r,m).
 \label{eq:A_lx_ss}
\end{equation}

\textit{Coarse grained learning rate} --
The coarse grained learning rate in Eq. \eqref{eq:coarsel2} becomes 
\begin{equation}
 l_\rr=-\int\dd x\int\dd r{J}_\rr(x,r)\frac{\del}{\del r}\ln{\rho}(x|r),
\end{equation}
where ${J}_\rr(x,r)\equiv\int\dd mJ_\rr(x,r,m)$, ${\rho}(x,r)\equiv\int\dd m\rho(x,r,m)$ and ${\rho}(x|r)\equiv {\rho}(x,r)/[\int{\rho}(\tilde x,r)\dd \tilde x]$.

\textit{Entropy production} -- The entropy production in \eqref{eq:sigma_x_master} is separated into two contributions 
\begin{equation}
 \sigma_\y\equiv \sigma_\rr+\sigma_\m,
\end{equation}
as shown in Eqs. \eqref{entr} and \eqref{entm}. In the continuous limit, using Eqs \eqref{eq:FP_LNA_pre} and \eqref{eq:A_current_def}, these contributions become
\begin{equation}
 \sigma_\rr=\int\dd x\int\dd r\int \dd m J_\rr(x,r) F_\rr(x,r),
 \label{eq:A_sigma_r}
\end{equation}
and
\begin{equation}
 \sigma_\m=\int\dd x\int\dd r\int \dd m J_\m (x,r,m) F_\m(r,m).
 \label{eq:A_sigma_m}
\end{equation}

\textit{Coarse grained entropy production} -- From Eqs. \eqref{eq:FP_LNA_pre}, \eqref{eq:A_current_def} and \eqref{eq:A_sigma_m}, the coarse grained entropy production \eqref{eq:sigma_cg}
becomes
\begin{multline}
\tilde{\sigma}_\y=\\
\int\dd r\int\dd m\left[\int J_\rr(x,r,m) \dd x\right]\left[\int F_\rr(\tilde x,r)\rho(\tilde x|r,m) \dd \tilde x\right]\\
+\sigma_\m
\label{eq:A_sigma_cg}
\end{multline}
The last term $\sigma_\m$ remains the same  because $m$ is not directly influenced by the signal $x$.

\textit{Upper bound on transfer entropy rate} -- 
The upper bound of the transfer entropy rate \eqref{eq:Tsxu_master} becomes
\begin{multline}
 \quad\overline{\mathcal{T}}_{\x\to\y}=
\\
 \frac{D_\rr}{4}\int \dd x\int\dd r\int\dd m\rho(x,r,m)\left[F_\rr(x,r)^2-\tilde{F}_\rr(r,m)^2\right],\hidewidth\\
 \label{eq:A_Tsxu}
\end{multline}
where we used the averaged force
\begin{equation}
 \tilde{F}_\rr(r,m)\equiv\int\dd x\rho(x|r,m) F_\rr(x,r).
 \label{eq:A_Tsxu_aux_2}
\end{equation}
Since, $\tilde{F}_\m(r,m)=F_\m(r,m)$ the  contribution due to $m$ is zero.
For $\overline{\mathcal T}_{\x\to\rr}$  defined in Eq. \eqref{eq:uppertransfer_def2} we replace $\rho(x|r,m)$ by $\rho(x|r)$ in Eqs. \eqref{eq:A_Tsxu_aux_2} and \eqref{eq:A_Tsxu}, which leads to the expression  
\begin{equation}
 \quad\overline{\mathcal{T}}_{\x\to\rr}= \frac{D_\rr}{4}\int \dd x\int\dd r\rho(x,r)\left[F_\rr(x,r)^2-\tilde{F}_\rr(r)^2\right],\hidewidth\\
 \label{eq:A_Tsru}
\end{equation}
where $\tilde{F}_\rr(r)\equiv\int\dd x\rho(x|r) F_\rr(x,r)$.

\subsection{Gaussian linear processes}

\label{subsec:A_gaussian}
We now consider a linear Langevin equation of the form
\begin{equation}
 \begin{pmatrix}
  \dot x_t\\
  \dot \by_t
 \end{pmatrix}
=-\mat{A}\begin{pmatrix}
   x_t\\
   \by_t
 \end{pmatrix}
 +\boldsymbol{\xi}_t,
 \label{eqeqeq}
 \end{equation}
where $\avg{\boldsymbol{\xi}_t\boldsymbol{\xi}_t^\T}=2\mat{D}\delta(t-t')$.
The matrices $\mat A$ and $\mat D$ for the bare sensor $\by=r$ are given by
\eqref{eq:2x2matrixAD}  and for the sensor with a memory $\by=(r,m)$ they are given by \eqref{eq:3x3matrixAD}.
The steady state solution of this Langevin equation is a multivariate normal distribution $\rho(x,\by)$ with zero mean and covariance $\bSigma$, which is the stationary solution of \eqref{eq:lypunov}.
Comparing Eqs. \eqref{langevingen} and \eqref{eqeqeq} the drift term is
\begin{equation}
\boldsymbol{F}(x,\by)\equiv-
\mat {D}^{-1}\mat{A}\begin{pmatrix}
x\\
\by \end{pmatrix}.
\label{eq:A_v_gauss}
\end{equation}
The probability current defined in Eq. \eqref{eq:A_current_def} is then given by
\begin{equation}
 \boldsymbol{J}(x,\by)=-\left[\mat{A}
 -\mat{D}\bSigma^{-1}\right]\begin{pmatrix}
   x\\
   \by
 \end{pmatrix}\rho(x,\by),
 \label{eq:A_J_gauss}
\end{equation}
where $\bSigma^{-1}$ is the inverse of $\bSigma$.

We define the matrix
\begin{equation}
 \boldsymbol{\Phi}\equiv\int \dd x\int \dd\by
\boldsymbol{J}(x,\by)\boldsymbol{F}(x,\by)^\T.
\end{equation}
Eqs. \eqref{eq:A_v_gauss} and \eqref{eq:A_J_gauss} yield
\begin{equation}
 \boldsymbol{\Phi}=\left[\mat{A} 
 -\mat{D}\bSigma^{-1}\right]\bSigma\mat{A}^\T\mat{D}^{-1}=\mat{A}\bSigma\mat{A}^\T\mat{D}^{-1} -\mat{D}\mat{A}^\T\mat{D}^{-1},
\end{equation}
where we used the fact that $\rho(x,\by)$ is a multivariate Gaussian density.
With this expression, from Eq. \eqref{eq:A_sigma_r} we obtain 
\begin{equation}
 \sigma_\rr=\Phi_{\rr\rr}=\omega_\x\frac{\nu_\rr^2}{B_\rr(1+\nu_\rr)},
\label{eqappsigmar}
\end{equation}
and from Eq. \eqref{eq:A_sigma_m} we obtain
\begin{equation}
 \sigma_\m=\Phi_{\m\m}=\omega_\x\frac{\nu_\m^2[\nu_\rr^2+B_\rr(1+\nu_\m)(1+\nu_\rr)]}{B_\m (1+\nu_\m)(1+\nu_\rr)(\nu_\m+\nu_\rr)},
 \label{eq:A_sigma_m_linear}
\end{equation}
where $\mathcal{E}_\x^2\equiv D_\x/\omega_\x,\nu_\rr\equiv \omega_\rr/\omega_\x,B_\rr \equiv D_\rr/D_\x,\nu_\m\equiv \omega_\m/\omega_\x,B_\m\equiv D_\m/D_\x$ (as defined in Sec. \ref{sec:linear}). 

The gradient of the log of the density reads
\begin{equation}
 \boldsymbol{a}(x,\by)\equiv
 -
 \begin{pmatrix}
 \del_x\\\del\by
 \end{pmatrix}
 \ln\rho(x,\by)
 =\bSigma^{-1}
 \begin{pmatrix}
  x\\\by
 \end{pmatrix}.
 \label{eq:A_help1}
\end{equation}
With the matrix
\begin{align}
 \mat{L} & \equiv\int\dd x\int\dd \by\boldsymbol{J}(x,\by)\boldsymbol{a}(x,\by)^\T\nonumber\\
 & = -(\mat{A}-\mat{D}\bSigma^{-1})\bSigma\bSigma^{-1}=-\mat{A}+\mat{D}\bSigma^{-1},
\end{align}
where we used Eqs. \eqref{eq:A_J_gauss}  and \eqref{eq:A_help1}, the learning rate $l_\y=\mat{L}_{\x\x}$ \eqref{eq:A_lx_ss} reads
\begin{equation}
 l_\y=\mat{L}_{\x\x}=\omega_\x\Big[-1+\mathcal{E}_\x^2(\bSigma^{-1})_{\x\x}\Big].
 \label{eq:A_lx_gauss_gen}
\end{equation}
The 2$\times$2 covariance matrix of $(x,r)$
given by \eqref{eq:covariance}
yields
\begin{equation}
 l_\rr=\mat{L}_{\x\x}=\omega_\x\frac{\nu_\rr^3}{\nu_\rr^2+B_\rr(1+\nu_\rr)^2}.
\end{equation}
For a the case with memory, where $(x,\by)=(x,r,m)$, the explicit form of the learning rate \eqref{eq:A_lx_gauss_gen} is given by
\begin{widetext}
 \begin{align}
  &l_\y=\mat{L}_{\x\x}
  =\nonumber\\&
  \!\!\!\!\!\!\!\!\!\!\!\!\!
\frac{\omega_\x\nu _\rr^2 \left(\nu _\m+\nu _\rr\right) \left\{B_\rr \nu _\m^2 \left(\nu _\m \nu _\rr+1\right)+\nu _\rr \left[B_\m \left(\nu _\m+1\right){}^2 \left(\nu _\m+\nu _\rr\right)+\nu _\m^2 \nu _\rr\right]\right\}}{
\nu _\m^2 \left\{B_\rr \nu _\rr^2 \left[\nu _\m^2+\nu _\m \left(4 \nu _\rr+2\right)+\nu _\rr^2 +2\nu _\rr+2\right]+B_\rr^2 \left(\nu _\m+1\right){}^2 \left(\nu _\rr+1\right){}^2+\nu _\rr^4\right\}+B_\m \left(\nu _\m+1\right){}^2
\left[B_\rr \left(\nu _\rr+1\right){}^2+\nu _\rr^2\right] \left(\nu _\m+\nu _\rr\right){}^2}.
\label{lyexpression}
\end{align}
\end{widetext}
The upper bound on the transfer entropy rate \eqref{eq:A_Tsxu} reads 
\begin{equation}
\overline{\mathcal T}_{\x\to \y} =\frac{\omega_\rr^2}{4 D_\rr}\int\dd x\int \dd r\int \dd m\,\rho(x,r,m)\left[x^2-\avg{x|r,m}^2\right],
\label{eq:A_Tsxu_langevin}
\end{equation}
where $\avg{x|r,m}\equiv\int\rho(\tilde x|r,m)\tilde x\,\dd\tilde x$ and we used $F_\rr(x,r)=\omega_\rr( x-r)/D_\rr$. 

\section{Uncertainty from instantaneous state and from time-series}
\label{sec:A_Un}

We first consider a sensor with memory $\by= (r,m)$. The covariance matrix, which is the stationary solution  of \eqref{eq:lypunov} with matrices given by \eqref{eq:3x3matrixAD}, is written as
\begin{equation}
 \bSigma=
 \begin{pmatrix}
  \varSigma_{\x\x}&\varSigma_{\x\rr}&\varSigma_{\x\m}\\
  \varSigma_{\x\rr}&\varSigma_{\rr\rr}&\varSigma_{\rr\m}\\
  \varSigma_{\x\m}&\varSigma_{\rr\m}&\varSigma_{\m\m}\\
 \end{pmatrix}
 \equiv
 \begin{pmatrix}
 \mathcal{E}_\x^2&\boldsymbol{b}^\T\\
 \boldsymbol{b}&\tilde{\bSigma}\\
 \end{pmatrix}.
 \label{covm}
\end{equation}
The linear estimate of $x$ from $\by$ is $\hat{x}(\by)\equiv\bc^\T\by$, where $\mat{c}$  is a vector.
Minimizing the variance 
\begin{equation}
\avg{[x-\hat x(\by)]^2}=\mathcal{E}_\x^2-2\bc^\T\bbb+\bc^\T\tilde{\bSigma}\bc,
 \label{eq:A_Exy_pre}
\end{equation}
which is minimal for $\bc=\tilde{\bSigma}^{-1}\bbb$, leads to the uncertainty 
\begin{equation}
 \mathcal{E}_{\x|\y}^2=\mathcal{E}_\x^2-\bbb^\T\tilde{\bSigma}^{-1}\bbb=\mathcal{E}_\x^2\left(1-\frac{\bbb^\T\tilde{\bSigma}^{-1}\bbb}{\mathcal{E}_\x^2}\right).
 \label{eq:A_Exy_final}
\end{equation}
Following the same procedure for a bare sensor with $\by= r$, $\tilde{\bSigma}=\varSigma_{\rr\rr}$, and $\bbb=\varSigma_{\x\rr}=\bbb^\T$ the covariance matrix \eqref{eq:covariance} leads to an uncertainty  
\begin{equation}
\mathcal{E}_{\x|\rr}^2= \mathcal{E}_\x^2\left[1-\frac{\nu_\rr^3}{\nu_\rr^3+\nu_\rr^2+B_\rr(1+\nu_\rr)^2}\right].
\label{eq:error_state}
\end{equation}
Comparing Eq. \eqref{eq:uppertransfer_example} with Eq. \eqref{eq:error_state} we obtain 
\begin{equation}
\overline{\mathcal{T}}_{\x\to \rr}=\frac{\omega_\x\nu_\rr^2}{4B_\rr}\frac{\mathcal{E}_{\x|\rr}^2}{\mathcal{E}_{\x}^2}.
\label{transfer1}
\end{equation}
Likewise, from Eq. \eqref{eq:A_Tsxu_langevin}, with $\rho(x,r,m)$ a multi-variative Gaussian with zero mean and covariance matrix \eqref{covm}, and Eq. \eqref{eq:A_Exy_final} we obtain 
\begin{equation}
\overline{\mathcal{T}}_{\x\to \y}=\frac{\omega_\x\nu_\rr^2}{4B_\rr}\frac{\mathcal{E}_{\x|\y}^2}{\mathcal{E}_{\x}^2}.
\label{transfer2}
\end{equation}

The best estimate $\hat{x}_t$ that uses the time-series of the sensor $\{r_{t'}\}_{t'\le t}$ to minimize the uncertainty $\hat{\mathcal{E}}_t^2\equiv\avg{(x_t-\hat{x}_t)^2}$ is known
as the Kalman-Bucy filter \cite{okse03,horo14a}. For the linear Gaussian process from \eqref{langevin3} 
the best estimate $\hat{x}_t$ satisfies $\avg{r_{t'}\hat{x}_{t}}=\avg{r_{t'}x_t}$ for all $t'\le t$ and $\avg{\hat{x}_t(x_t-\hat{x}_t)}=0$ (see \cite{okse03}).
It can be shown that the minimal error satisfies the Riccati equation, which reads \cite{okse03,horo14a}  
\begin{equation}
 \frac{\dd}{\dd t}\hat{\mathcal{E}}_t^2=-\frac{\omega_\rr^2}{2D_\rr}\hat{\mathcal{E}}_t^4-2\omega_\x\hat{\mathcal{E}}_t^2+2D_\x.
 \label{eq:A:Riccati}
\end{equation}
The stationary solution of this equation gives the uncertainty about the signal given the sensor trajectory 
\begin{equation}
 \mathcal{E}_{\x|\rr_\text{traj}}^2=\mathcal{E}_\x^2\left(\frac{2}{1+\sqrt{1+\frac{\nu_\rr^2}{B_\rr}}}\right).
\end{equation}
Comparing with Eq. \eqref{eq:transfer_example} we obtain
\begin{equation}
\mathcal{T}_{\x\to \rr}=\frac{\omega_\x\nu_\rr^2}{4B_\rr}\frac{\mathcal{E}_{\x|\rr_\text{traj}}^2}{\mathcal{E}_{\x}^2}.
\label{transfer3}
\end{equation}
The simple relations \eqref{transfer1}, \eqref{transfer2}, and \eqref{transfer3}  are valid for our model system that corresponds 
to a linear Gaussian process. Since for $C=1$ the transfer entropy rate equals its upper bound, for our model system a maximal 
sensory capacity $C=1$ implies $\mathcal{E}_{\x|\rr_\text{traj}}=\mathcal{E}_{\x|\y}$.
In this case the linear estimate $\hat x (\by)=\bc^\T\by=\bbb^\T\tilde{\bSigma}^{-1}\by$ from Eq. \eqref{eq:A_Exy_pre} coincides with the estimate 
from the Kalman-Bucy filter $\hat{x}_t$, which is similar to the finding in \cite{horo14a} for optimal feedback cooling.

\bibliography{sensory_capacity}


 \end{document}